\documentclass[preprint,3p]{elsarticle}
\pdfoutput=1

\usepackage{amsmath}
\usepackage{amssymb}
\usepackage{graphicx}
\usepackage{bbding}
\usepackage{array}
\usepackage{color}
\usepackage[T1]{fontenc} 
\usepackage{slashed}
\usepackage{xspace}
\usepackage{ulem}
\usepackage{listings}
\usepackage{xcolor}
\usepackage{color}
\newcommand\SARAH{{\tt SARAH}\xspace}
\newcommand\SPheno{{\tt SPheno}\xspace}
\newcommand\FlexibleSUSY{{\tt FlexibleSUSY}\xspace}
\newcommand\SoftSUSY{{\tt SOFTSUSY}\xspace}
\newcommand\NMSSMCalc{{\tt NMSSMCALC}\xspace}
\newcommand\NMSSMTools{{\tt NMSSMTools}\xspace}

\newcommand\Mathematica{{\tt Mathematica}\xspace}

\newcommand\FlexibleDecay{{\tt FlexibleDecay}\xspace}
\newcommand\NMSPEC{{\tt NMSPEC}\xspace}
\newcommand\NMHDECAY{{\tt NMHDECAY}\xspace}
\newcommand\NMSDECAY{{\tt NMSDECAY}\xspace}
\newcommand\NMGMSB{{\tt NMGMSB}\xspace}
\newcommand\MO{{\tt MicrOmegas}\xspace}
\newcommand\HDECAY{{\tt HDECAY}\xspace}

\newcommand{\yes}{\Checkmark}
\newcommand{\no}{\XSolidBrush}
\newcommand{\DR}{\ensuremath{\overline{\text{DR}}}\xspace}
\newcommand{\MS}{\ensuremath{\overline{\text{MS}}}}
\newcommand{\gev}{\text{GeV}}
\newcommand{\re}{\Re\text{e}}

\begin{document}
\title{Higgs mass predictions of public NMSSM spectrum generators}
%
%
%

\author[fs]{Florian Staub\corref{cor1}}
\address[fs]{Theory Division, CERN, 1211 Geneva 23, Switzerland}
\ead[fs]{florian.staub@cern.ch}
\author[pa]{Peter Athron}
\address[pa]{ARC
Centre of Excellence for Particle Physics at the Terascale, School of Physics, Monash University, Melbourne VIC 3800, Australia }
\ead[pa]{peter.athron@coepp.org.au}
\author[ue]{Ulrich Ellwanger}
\address[ue]{LPT, UMR 8627, CNRS, Universit\'e de Paris-Sud, 91405 Orsay, France, and
School of Physics and Astronomy, University of Southampton,
Highfield, Southampton SO17 1BJ, UK}
\ead[ue]{Ulrich.Ellwanger@th.u-psud.fr}
\author[rg]{Ramona Gr\"ober}
\address[rg]{Sezione di Roma Tre, Via della Vasca Navale 84, I-00146 Roma, Italy}
\ead[rg]{groeber@roma3.infn.it}
\author[mm]{Margarete M\"uhlleitner}
\address[mm]{Institute for Theoretical Physics, Karlsruhe Institute of Technology,
Wolfgang-Gaede-Str. 1, 76131 Karlsruhe, Germany}
\ead[mm]{maggie@particle.uni-karlsruhe.de}
\author[ps]{Pietro Slavich}
\address[ps]{LPTHE,  UPMC Univ. Paris 06, Sorbonne Universit\'es,  UMR 7589, 4 Place Jussieu,  F-75252, Paris, France\\
 LPTHE, CNRS, UMR 7589, 4 Place Jussieu, F-75252, Paris, France}
 \ead[ps]{slavich@lpthe.jussieu.fr}
\author[av]{Alexander Voigt}
\address[av]{Deutsches Elektronen-Synchrotron DESY, 
Notkestra\ss e 85, D-22607 Hamburg, Germany}
\ead[av]{alexander.voigt@desy.de}
 \cortext[cor1]{Corresponding author}
\date{\\CERN-PH-TH-2015-164, DESY 15-122, KA-TP-16-2015, LPT-Orsay-15-54, RM3-TH/15-11}

\begin{abstract}
  The publicly available spectrum generators for the NMSSM often lead
  to different predictions for the mass of the standard model-like
  Higgs boson, even though they all implement two-loop
    computations in the \DR renormalization scheme.  Depending on the
  parameter point, the differences can exceed 5~GeV, and even reach
  8~GeV for moderate superparticle masses of up to 2~TeV. It is shown
  here that these differences can be traced
  back to the calculation of the running standard model parameters
  entering all calculations, to the approximations used in the
  two-loop corrections included in the different codes, and to
  different choices for the renormalization conditions and scales. In
  particular, the importance of the calculation of the top Yukawa
  coupling is pointed out.
\end{abstract}
\maketitle

\section{Introduction}

Now that run II of the Large Hadron Collider (LHC) has started, either
the first clear hints for new physics beyond the standard model (BSM)
will finally show up or many well-motivated theories for BSM physics
will come under even greater pressure. While in run I a scalar was
discovered \cite{Chatrchyan:2012ufa,Aad:2012tfa} which has all
properties of the long-expected Higgs boson of the standard model
(SM), at the same time stringent limits were placed on the simplest
scenarios for new physics. In particular, the minimal realization of
supersymmetry -- the minimal supersymmetric standard model (MSSM) --
has lost at least some of its appeal: no hints for supersymmetric
(SUSY) particles were found, and the 
observed value of the SM-like Higgs mass
requires either multi-TeV stop masses or special values of the
  stop-mixing parameter.
However, if the stops are heavy and the MSSM is assumed to be a
  valid description of nature far above the electroweak scale, then a
  large fine-tuning is required to stabilize
  the vacuum expectation value (VEV) of the Higgs field. On
  the other hand, the large stop mixing needed to generate a Higgs mass of $125$~GeV for 
  stop masses around one
  TeV turns out to be dangerous because of charge and color breaking
minima
\cite{Camargo-Molina:2013sta,Blinov:2013fta,Chowdhury:2013dka,Camargo-Molina:2014pwa,Chattopadhyay:2014gfa}. This
leads to the question of whether the MSSM is really a natural
completion of the SM and has attracted more interest in other SUSY
scenarios. This is especially the case
for SUSY models that
  predict a larger Higgs mass already at tree level. The best-studied
examples are models with singlet extensions, where the
interaction between the Higgs fields and the gauge singlet can raise
the Higgs mass \cite{Maniatis:2009re,Ellwanger:2009dp,Ellwanger:2006rm} and
significantly reduce the
fine-tuning~\cite{BasteroGil:2000bw,Dermisek:2005gg,Ellwanger:2011mu,Ross:2011xv,Ross:2012nr,Gherghetta:2012gb,Perelstein:2012qg,Kim:2013uxa,Kaminska:2014wia,Binjonaid:2014oga}.
A $\mathbb{Z}_3$ symmetry is often introduced to forbid all
dimensionful parameters in the superpotential and to solve the
$\mu$~problem of the MSSM at the same time
\cite{Ellwanger:2009dp}. This realization of a singlet extension of
the MSSM is the next-to-minimal supersymmetric standard model
(NMSSM). The NMSSM comes with a rich collider phenomenology
\cite{Dreiner:2012ec,King:2012tr,King:2014xwa}, and also provides
possibilities to hide SUSY at the LHC \cite{Ellwanger:2014hia}.

To confront the NMSSM with the measured Higgs mass, a precise
calculation is necessary.  The first calculations of radiative
corrections to the Higgs masses in the NMSSM assumed an effective
field theory below the SUSY scale (the scale of the stop/sbottom
masses), and integrated the renormalization group equations (RGEs) for
the corresponding couplings from the SUSY scale to the weak scale
\cite{Espinosa:1991fc,Espinosa:1992hp,Elliott:1993ex}.  One-loop
corrections\footnote{We define $\alpha_{t,b,\tau} =
  Y_{t,b,\tau}^2/(4\pi)$, where $Y_{t,b,\tau}$ are the third-family
  Yukawa couplings, and adopt analogous definitions for
    $\alpha_\lambda$ and $\alpha_\kappa$, where $\lambda$ and $\kappa$
    are the NMSSM-specific couplings in eq.~(\ref{eq:GNMSSM})
    below. We follow the standard convention of denoting, e.g., by
  ${\cal O}(\alpha_t)$ the one-loop corrections to the Higgs masses
  that are in fact proportional to $\alpha_t \,m_t^2$,
  i.e.~proportional to $Y_t^4$. Similar abuses of notation affect the
  other one- and two-loop corrections.} of ${\cal O}(\alpha_t +
\alpha_b)$ in the effective potential approximation have been computed
in
Refs.~\cite{Ellwanger:1993hn,Pandita:1993tg,Elliott:1993uc,Elliott:1993bs}.
Two-loop corrections of ${\cal O}(\alpha_s\, \alpha_t +
  \alpha_t^2)$ have been included in the NMSSM in the
leading-logarithmic approximation (LLA) in
Refs.~\cite{Yeghian:1999kr,Ellwanger:1999ji}, and one-loop corrections
to the SM-like Higgs mass including NMSSM-specific and electroweak
couplings in the LLA in Ref.~\cite{Ellwanger:2005fh}.  Full one-loop
calculations including the momentum dependence have been performed in
the $\overline{\mathrm{DR}}$ renormalization scheme\footnote{Strictly
  speaking, all calculations have been done in the
  $\overline{\mathrm{DR}}'$-scheme where the $\epsilon$ scalars are
  decoupled
  \cite{Martin:2001vx}.}~\cite{Degrassi:2009yq,Staub:2010ty}, or in a
mixed on-shell (OS)-\DR\ scheme~\cite{Ender:2011qh,Graf:2012hh}; at
the two-loop level, the dominant ${\cal
  O}(\alpha_s(\alpha_b+\alpha_t))$ corrections in the
$\overline{\mathrm{DR}}$ scheme have become available a few years ago
\cite{Degrassi:2009yq}. The two-loop corrections involving only
superpotential couplings such as Yukawa and singlet
interactions~\cite{Goodsell:2014pla}, and a two-loop calculation of
the ${\cal O}(\alpha_s \alpha_t)$ corrections with the top/stop sector
renormalized either in the OS scheme or in the
$\overline{\mathrm{DR}}$ scheme~\cite{Muhlleitner:2014vsa}, have also
been provided recently.

At the moment, five public spectrum generators for the NMSSM are
available which make use of these results to compute the Higgs-boson
masses and mixing matrices in the NMSSM for a given parameter point:
the stand-alone codes \NMSSMTools
\cite{Ellwanger:2004xm,Ellwanger:2005dv,Ellwanger:2006rn}, \SoftSUSY
\cite{Allanach:2001kg,Allanach:2013kza,Allanach:2014nba} and
\NMSSMCalc \cite{Baglio:2013iia}, as well as NMSSM versions of
\FlexibleSUSY \cite{Athron:2014yba} and \SPheno
\cite{Porod:2003um,Porod:2011nf}, which are based on \SARAH
\cite{Staub:2008uz,Staub:2009bi,Staub:2010jh,Staub:2012pb,Staub:2013tta}.
The various tools, however, often make predictions that differ by
several GeV for the mass of the SM-like Higgs boson even if using the
same renormalization scheme.  These differences are bigger than
  the ones observed between codes that calculate
  the Higgs mass in the MSSM \cite{Allanach:2004rh,Allanach:2003jw}.
The aim of this work is to identify the origin of these deviations for
the \DR calculation, and to determine the numerical significance of
each difference between the codes. A comparison 
  between the \DR and OS calculations, 
  which would provide an estimate of the
  theoretical uncertainty, is beyond the scope of this
  analysis and will be given elsewhere.

The paper is organized as follows. In sec.~\ref{sec:conventions} we
introduce the model and the conventions used. In sec.~\ref{sec:codes}
the different codes and their features are presented, and the results
obtained in six test points when using the codes out-of-the
box are shown. The differences in the Higgs-mass calculation between
these codes are discussed in sec.~\ref{sec:differences}. Here, we
concentrate on a SUSY scale input, while we address additional effects
in a GUT scenario in sec.~\ref{sec:GUT}. We conclude in
sec.~\ref{sec:conclusion}.

\section{The NMSSM and its conventions}
\label{sec:conventions}

We consider in the following a model with the particle content of the
MSSM extended by a gauge singlet superfield $\hat S$. The general,
renormalizable and $R$-parity conserving superpotential for this model
in terms of the (hatted) doublet ($\hat{H}_u, \hat{H}_d$) and singlet superfields reads
\begin{equation}
\label{eq:GNMSSM}
 \mathcal{W}_S = \mathcal{W}_\text{Yukawa}  + \frac{1}{3}\kappa \hat S^3+ (\mu + \lambda \hat S) \hat H_u \hat H_d +  \frac{1}{2} \mu_s \hat S^2 + t_s \hat S  \;,
\end{equation}
with 
\begin{equation}
\label{eq:Yukawa}
\mathcal{W}_\text{Yukawa}  = Y_u \hat Q \hat U \hat H_u + Y_d \hat Q \hat D \hat H_d + Y_e \hat L \hat E \hat H_d \;,
\end{equation}
where we have suppressed color and isospin indices in every term. In
general, the couplings in eq.~(\ref{eq:GNMSSM}) are complex
parameters, and those in eq.~(\ref{eq:Yukawa}) are complex $3\! \times\!
3$ matrices. The most general soft SUSY-breaking terms for this theory
are given by
\begin{align}
 V_\text{soft} 
   &=  m_s^2 |S|^2 + m_{H_u}^2 |H_u|^2+ m_{H_d}^2 |H_d|^2 + \tilde{f}^\dagger m_{\tilde{f}}^2 \tilde{f} \nonumber \\
   & + M_1 \lambda_B^2 + M_2 \lambda_W^2 + M_3 \lambda_g^2 + T_d \tilde{q} \tilde{d} H_d + T_u \tilde{q} \tilde{u} H_u + T_e \tilde{e} \tilde{l} H_d \nonumber \\
   &+ \left(B_\mu H_u H_d + \frac12 B_s S S + \chi_s S +  T_\lambda S H_u H_d + \frac{1}{3} T_\kappa S^3  + h.c.\right) \;,
\label{soft}
\end{align}
with the left-handed squark and slepton doublets $\tilde{f}=\tilde q$
and $\tilde l$ and the right-handed squarks and sleptons
$\tilde{f}=\tilde{u}_R, \tilde{d}_R, \tilde{e}_R$. All generation indices are suppressed.
The soft
SUSY-breaking masses $m_s^2$, $m_{H_d}^2$ and $m_{H_u}^2$ are always
real, and the sfermion masses $m^2_{\tilde f}$ are hermitian
$3\!\times\! 3$ matrices, while all other soft terms can be complex. The
soft SUSY-breaking trilinears $T_\lambda$ and $T_\kappa$ are often
expressed in an alternative way, as the product of the corresponding
superpotential coupling times a free
parameter with the dimensions of a mass
\begin{eqnarray}
 & T_\lambda \equiv  \lambda A_\lambda \,,\hspace{2cm}  
T_\kappa \equiv  \kappa A_\kappa & \;.
\end{eqnarray}
Similarly, the trilinear soft SUSY-breaking interactions involving
sfermions can be expressed as
\begin{align}
  T_i \equiv & Y_i A_i \;, \hspace{1cm} i=u,d,e  \;.
\end{align}
It is common to introduce a $\mathbb{Z}_3$ symmetry to forbid all
dimensionful parameters (i.e., $\mu$, $\mu_s$ and $t_s$) in
eq.~(\ref{eq:GNMSSM}). This solves the $\mu$~problem of the MSSM
\cite{Kim:1983dt} -- at the price of introducing a domain wall problem
\cite{Abel:1995wk} -- and leads to the NMSSM, where the
surviving terms in the superpotential are
\begin{equation}
\label{eq:NMSSM}
\mathcal{W}_\text{NMSSM} = \mathcal{W}_\text{Yukawa}  + \frac{1}{3}\kappa \hat S^3+ \lambda \hat S \hat H_u \hat H_d \;,
\end{equation}
and the soft SUSY-breaking terms $B_\mu$, $B_s$ and $\chi_s$ in
eq.~(\ref{soft}) also vanish.

For suitable values of the remaining $s$-dependent SUSY-breaking terms $m_s$ and $T_\kappa$, a VEV $v_s$ for the scalar singlet $S$ of the order of the SUSY-breaking scale is generated.
 This dynamically
generates the $\mu$ term which gives masses to the higgsinos,
\begin{equation}
\mu_{\rm eff} = \lambda \langle S \rangle \;.
\end{equation}
Note, there are two common conventions: $\langle S \rangle = v_s$ or
$\langle S \rangle = v_s/\sqrt{2}$. The SUSY Les Houches Accord
  (SLHA)~\cite{Skands:2003cj,Allanach:2008qq} does not
  fix a convention, and the  codes
  compared here make different choices. However in this text we will
  use the latter convention. Finally, after electroweak symmetry
breaking (EWSB) where also the neutral components of $H_d$ and $H_u$
receive VEVs $v_d$ and $v_u$, the singlet mixes with the other
doublets.\footnote{While in the CP violating NMSSM the VEVs can be
  taken complex, it turns out that, after exploiting the minimum
  conditions of the scalar potential, the NMSSM Higgs sector at tree
  level can be described by one complex phase, which is given by
  $\varphi_y = \varphi_\kappa - \varphi_\lambda + 2 \varphi_s -
  \varphi_u$, where $\varphi_u$ and $\varphi_s$ are the complex phases
  of the VEVs of $H_u$ and $S$, respectively, and $\varphi_\kappa$ and
  $\varphi_\lambda$ the complex phases of $\kappa$ and $\lambda$.}  In
the CP-conserving case, this leads to three CP-even physical scalars
and two CP-odd ones together with the neutral would-be Goldstone
boson.  It is common to define the basis for the CP-even states as
$(\phi_d,\phi_u,\phi_s)$ where $\phi_i$ are the real parts of the
neutral components of $H_d$, $H_u$ and $S$. This leads to the
tree-level mass matrix $\left(M_h^2\right)^{\rm tree}$, derived from
the tree-level scalar potential
\begin{equation}
 \left(M_{h}^{2}\right)_{ij}^{\rm tree} 
~= ~\frac{~~\partial^2 V^{\rm tree}}{\partial \phi_i \partial \phi_j} \;.
\end{equation}
The pole masses $m_{h_i}$ ($i=1,2,3$),
identified with the masses of the three CP-even physical scalars, are
the eigenvalues of the momentum dependent, loop-corrected mass matrix
\begin{equation}
\label{mfull}
M_h^2(p^2) ~= ~  \left(M_{h}^{2}\right)^{\rm tree}
+~ \sum_n^\infty \Pi^{(n)} (p^2)\;,
\end{equation}
each evaluated at $p^2 = m_{h_i}^2$. The $\Pi^{(n)} (p^2)$ denote
$3\times 3$ matrices of $n$-loop self-energies calculated for an
external momentum $p^2$. As mentioned in the introduction, only the
one-loop contribution, $\Pi^{(1)} (p^2)$, is completely known so
far. For $\Pi ^{(2)}$ only partial results in the limit $p^2=0$ exist
in the NMSSM. Different choices of the renormalization scheme lead to
different results for the physical scalar masses if the perturbative
series in eq.~(\ref{mfull}) is truncated at a given order $n$. The
differences between the results are of ${\cal O}(\Pi^{(n+1)})$, and
provide an estimate of the theoretical uncertainty of the calculation
due to uncomputed higher-order corrections.  The matrix that
diagonalizes $M_h^2(p^2)$ is called $Z^H$ in the following\footnote{
 Note that, in eq.~(\ref{defZH}), only one
    element of the matrix $m_h^{diag}$ at a time can correspond to a
    pole Higgs mass, and only for an appropriate choice of $p^2$.  },
\begin{equation}
\label{defZH}
m_h^{diag} ~=~ Z_H \,M_h^2(p^2)\, Z_H^T  \;.
\end{equation}
In general, $Z_H$ is a $p^2$-dependent quantity. The NMSSM
spectrum generators adopt different definitions for the loop-corrected
rotation matrix $Z_H$ that they give as output: the external momentum
is either set equal to zero, or set equal to one of the mass
eigenvalues.

\bigskip

We conclude this section by presenting the set
of input parameters that enter the calculation of 
  the Higgs-boson masses and mixing matrices in the CP-conserving
  NMSSM. First,
there are  the three tadpole
conditions that ensure the minimum of the Higgs potential at
non-vanishing VEVs $v_u, v_d, v_s$,
\begin{equation}
\left.\frac{\partial V^{\rm eff}}{\partial \phi_i}\right|_{\text{min}}\!=~0 \;, \hspace{1cm} i=d,u,s \;,
\end{equation}
where $V^{\rm eff}$ represents the effective scalar potential computed
at the same perturbative order as the mass matrix in eq.~(\ref{mfull}).
These conditions fix three parameters.  The simplest choice is to
solve them for $m_s^2$, $m_{H_d}^2$ and $m_{H_u}^2$, and this is done
in all the comparisons for every code we use. Since one
combination of VEVs (namely, $v^2 = v_u^2 + v_d^2$) is constrained by
experimentally measured observables, this leaves six
free parameters in the Higgs sector of the model:
\begin{eqnarray}
  \lambda \,,\ \kappa \,,\ A_\lambda \,,\ A_\kappa \,,\ \tan\beta \,,\ \mu_{\rm eff} \;, 
\end{eqnarray}
where $\tan\beta \equiv {v_u}/{v_d}\,$, and $\mu_{\rm eff}$ stands in for
  $v_s$.  Under the assumption that the soft SUSY-breaking terms are flavor-diagonal and that only
third-generation soft interactions are non-negligible, this reduces
the additional input parameters required to fully define the
model to the set
\begin{eqnarray}
   A_t \,,\ A_b \,,\ A_\tau \,,\ m_{\tilde t_L} \,,\ m_{\tilde t_R} \,,\ M_1 \,,\ M_2 \,,\ M_3 \,,\ M_{\tilde f} \,,
\end{eqnarray}
with $m^2_{\tilde t_L} \equiv m_{\tilde q,33}^2$, $m^2_{\tilde t_R}
\equiv m_{\tilde u,33}^2$\footnote{For further reference, we define also
$m^2_{\tilde{u}_R} \equiv m^2_{\tilde{u}_{11}}$, $m^2_{\tilde{d}_R} \equiv m^2_{\tilde{d}_{11}}$, $m^2_{\tilde{u}_L} \equiv m^2_{\tilde{q}_{11}}$.}, $A_\tau = A_{e,33}$, $A_b = A_{d,33}$, $A_t = A_{u,33}$, and where we assumed for simplicity that
all other diagonal entries of $m_{\tilde f}^2$ are degenerate
and equal to $M^2_{\tilde f}$ . \\

One can further reduce the number of input parameters by assuming a
specific mechanism to transmit SUSY breaking from a hidden
sector to our visible sector. In models inspired by minimal supergravity the following relations among
the soft terms exist at the gauge-coupling unification scale:
\begin{equation}
m_{\tilde f}^2 ~=~ m^2_0 \,{\bf 1} \;,\hspace{5mm}  
M_i ~=~ M_{1/2}~~(i=1,2,3) \;, \hspace{5mm} 
A_j ~=~ A_0 ~~ (j=e,d,u)\;,
\end{equation}
where {\bf 1} is the
  identity matrix in flavor space. In contrast to the fully constrained NMSSM, the Higgs and singlet soft masses are obtained again by the tadpole equations, 
and also the NMSSM-specific trilinear soft couplings are treated as free parameters. Thus, necessary input parameters are
\begin{eqnarray}
\label{cnmssm}
m_0,\ M_{1/2},\ \tan\beta,\ A_0,\ \lambda,\ \kappa,\ A_\kappa,\, A_\lambda,\, \mu_{eff} \; .
\end{eqnarray}

\section{Spectrum generators for the NMSSM and their features}
\label{sec:codes}
\subsection{Features}
\begin{table}[htbp]
\begin{center}
\begin{tabular}{|p{2.25cm}| >{\centering\arraybackslash}p{2.25cm} >{\centering\arraybackslash}p{2.25cm}  >{\centering\arraybackslash}p{2.25cm} >{\centering\arraybackslash}p{2.25cm} >{\centering\arraybackslash}p{2.25cm}  |}
\hline 
                    &  \FlexibleSUSY               & \NMSSMCalc   & \NMSSMTools                  &  \SoftSUSY                      &   \SPheno \\
\hline      
\hline
\multicolumn{6}{|c|}{Code} \\
\hline
type                & using \SARAH              & stand alone & stand alone & stand alone &  using \SARAH \\[3mm]
language            & C++                          & Fortran 77 and 90               & Fortran 77                            & C++       & Fortran 90 \\
\hline
\multicolumn{6}{|c|}{Models} \\
\hline
No  $\mathbb{Z}_3$     &     \yes                            &  \no                &    \yes                              &  \yes                                & \yes \\
GUT model           &     \yes                            &  \no                &    \yes                              &  \yes                                & \yes \\
CPV                 &    (\yes)                              &   \yes              &   \no                             &  \no                                 & (\yes) \\
\hline   
\multicolumn{6}{|c|}{Thresholds} \\
\hline
scale(s) & $M_Z$ & $M_t\,, M_{SUSY}$ & $M_t\,, M_{SUSY}$ & $M_Z$ & $M_Z$  \\[5mm]
EW parameters & full one-loop      & OS definitions & full one-loop &  full one-loop  & full one-loop  \\[5mm]       
Yukawas  & full one-loop; two-loop QCD        & one-loop (S)QCD; two-loop QCD & one-loop (S)QCD+Yukawa; two-loop QCD     &  full one-loop; two-loop QCD; optionally two-loop SQCD   &  full one-loop   two-loop QCD   \\[5mm]     
strong gauge  & one-loop top+SUSY & --- & 
--- &  one-loop top+SUSY & one-loop top+SUSY   \\     
\hline
\multicolumn{6}{|c|}{Higgs mass calculation} \\
\hline
scheme              & \DR                                 & OS, \DR             & \DR                                  & \DR                                  & \DR \\[2mm]
one-loop              & full                              & full     & full                      & full                   & full\\[2mm]
two-loop              &$\alpha_s(\alpha_b+\alpha_t)$ +~MSSM & $\alpha_s \alpha_t$ & $\alpha_s(\alpha_b+\alpha_t)$ +~MSSM & $\alpha_s(\alpha_b+\alpha_t)$ +~MSSM & 
$\alpha_s \alpha_i + \alpha_i \alpha_j$.
\\
\hline 
\multicolumn{6}{|c|}{SUSY masses} \\
\hline
one-loop            &  \yes          & \no    & \yes & \yes & \yes \\
momentum effects    &  \yes          & \no    & \yes ($\alpha_s, \alpha_t, \alpha_b$ only)  & \yes & \yes \\
\hline
\multicolumn{6}{|c|}{Other observables} \\
\hline
decays              &    \no                             &   \yes  &   via {\tt NMHDECAY}                                         &  via {\tt NMHDECAY}                        & \yes \\            
flavour, $g-2$            & \no                                  & \no                            & \yes & \no & \yes  \\
\hline
\end{tabular}
\end{center}
\caption{Public spectrum generators for the NMSSM and their features. The first two rows specify if the code is stand alone or gets necessary information from the \Mathematica package \SARAH, and in what language the code is written. The next three rows list the models that are supported: ``GUT model'' means that the code accepts as input a set of SUSY-breaking parameters at the GUT scale, and ``CPV'' shows if a Higgs sector with CP-even and -odd mixing is supported. Here, (\yes) is used if this is only possible at the one-loop level so far.
  The rows under the sub-heading ``Thresholds'' give details of what SUSY and non-SUSY threshold corrections are included to calculate the running parameters entering the Higgs-mass calculation. The first row contains the scale (or scales) at which the threshold corrections are introduced, and the following three rows refer to different categories of parameters. 
  Under ``Higgs mass calculation'' the main aspects of the calculation of the Higgs mass are given: the renormalization scheme (``scheme''), and  what corrections at one- and two-loop are included. ``full'' at one-loop refers to a diagrammatic calculation including the entire $p^2$ dependence and all corrections up to those suppressed by first and second generation Yukawa couplings (which are also included in some codes). At two-loop the included corrections are shown.  ``+MSSM'' refers to the approximation that the known MSSM results for the pure Yukawa interactions at two-loop are included. In the last column, $\alpha_i$ and $\alpha_j$ stand for any of the couplings $\alpha_t$, $\alpha_b$, $\alpha_\tau$, $\alpha_\lambda$ and $\alpha_\kappa$. Under ``SUSY masses'' it is shown if the code calculates also one-loop corrections to SUSY states (neutralinos, charginos, sfermions, gluino) and if the effects from the external momenta are included in these calculations. Finally, it is summarized in ``Other observables'' if the tool gives also a prediction for flavor observables and/or other precision observables like $g-2$ and if it calculates the decays of at least the Higgs bosons. }
\label{tab:codes}
\end{table}
The main features of the publicly available spectrum generators for
the NMSSM are summarized in table~\ref{tab:codes}. Here, two different
kinds of spectrum generators are distinguished: dedicated tools which
contain all physically relevant information about the NMSSM (vertices,
mass matrices, renormalization group equations (RGEs), loop
corrections, etc.) in a hardcoded way, and multi-purpose generators
which use the \Mathematica package \SARAH to derive this information
for a given model. The advantage of the first approach, which is
 adopted by \NMSSMTools, \NMSSMCalc and \SoftSUSY,
is that the tools can be used directly after compilation without
 further effort. In contrast, to use
\FlexibleSUSY and \SPheno, 
one needs to first install \SARAH \footnote{\FlexibleSUSY also
  offers pre-generated code for selected models, with specific
  boundary conditions and EWSB outputs, allowing one to use these
  specific \FlexibleSUSY spectrum generators without first installing
  \Mathematica and \SARAH.}, which is then used to generate the
necessary source code to get an NMSSM spectrum generator. These
additional steps allow these tools to be used for a much broader range
of models, i.e.~not only other singlet extensions beyond the NMSSM are
supported, but also many other extensions of the MSSM. Also, the
features of the NMSSM version (like boundary conditions at all
relevant scales) can be adjusted in the input files for \SARAH or
\FlexibleSUSY without the need to modify the final code\footnote{For
  easier comparison with other codes, we have created input files to
  allow for purely SLHA1~\cite{Skands:2003cj} input for \SPheno and
  \FlexibleSUSY, while by default a block form for all matrices is
  used as in SLHA2 \cite{Allanach:2008qq}. Interested readers can
  obtain the files from the authors.}.  \SARAH is available at
hepforge
\begin{center}
    {\tt http://sarah.hepforge.org/} 
\end{center}

More details on the different tools are given in the following,
in the order in which the codes became publicly available. 

\subsubsection{\NMSSMTools} 
\label{sec:NMSSMTools}
The first public codes to be made available for the NMSSM were
\NMHDECAY~\cite{Ellwanger:2004xm,Ellwanger:2005dv},
\NMSPEC~\cite{Ellwanger:2006rn} and \NMGMSB, which are today
collected in the package
\NMSSMTools. \NMHDECAY was based on a suitably modified version of
{\tt HDECAY}~\cite{Djouadi:1997yw,Djouadi:2006bz,Butterworth:2010ym}.
This package is available at
 \begin{center}
  {\tt http://www.th.u-psud.fr/NMHDECAY/nmssmtools.html}
 \end{center}
 So far \NMSSMTools is restricted to the real NMSSM with and without
 $\mathbb{Z}_3$ symmetry, but a version to support also CP violation
 is under construction \cite{Domingo:2015qaa}. \NMSSMTools allows to
 calculate the Higgs masses for a parameter point with three different
 options:
\begin{enumerate}
\item The most precise calculation makes use of the NMSSM corrections
  of ref.~\cite{Degrassi:2009yq}. This provides a fully diagrammatic
  calculation of the Higgs masses at the one-loop level, including all
  contributions and the momentum dependence, plus
  the two-loop corrections of
  ${\cal O}(\alpha_s (\alpha_b + \alpha_t))$  at zero external
    momentum. In addition, the two-loop
  corrections known from the MSSM at  ${\cal
    O}((\alpha_t+\alpha_b)^2 + \alpha_\tau (\alpha_\tau +
  \alpha_b))$ \cite{Brignole:2001jy,Dedes:2002dy,Dedes:2003km,Allanach:2004rh}  are included.
\item The second option 
  implements the same corrections as the first one, but
  neglects the momentum dependence of the one-loop
    self-energies. (This makes the code faster, which is useful for
  large scans.)
\item The third option is the original implementation of the Higgs
  mass calculation in \NMSSMTools, which -- as described in the
    appendix C of ref.~\cite{Ellwanger:2009dp} -- implements only a
    partial computation of the one-loop contributions, and restricts
  the two-loop contributions to the leading-log approximation.
\end{enumerate}
\NMSSMTools also calculates the SUSY spectrum at the one-loop level,
and the Higgs and sparticle decay branching fractions (the latter in
{\tt NMSDECAY}~\cite{Das:2011dg} using a suitable modification of {\tt
  SDECAY}~\cite{Muhlleitner:2003vg}). $B$-physics observables and the
muon anomalous magnetic moment are computed following
refs.~\cite{Domingo:2007dx,Domingo:2008bb}, and compared to
experimental constraints.  Bounds on the Higgs sector from
LEP~\cite{Schael:2006cr} and on the couplings of the SM-like Higgs
boson (from ref.~\cite{Belanger:2013xza}) are implemented. A link to
\MO is included which allows for an easy calculation of the dark
matter relic density, and direct and indirect dark matter detection
cross sections.

It is possible to define a set of input
  parameters at the SUSY scale, or to define specific SUSY-breaking
scenarios such as minimal supergravity using \NMSPEC, or
scenarios of gauge-mediated SUSY breaking (following
ref.~\cite{Delgado:2007rz}) using \NMGMSB. In all cases, the
fine-tuning of the EWSB scale with respect to the input
parameters can be estimated.

\subsubsection{\SPheno and \SARAH} 
The main features of the \SPheno version generated with \SARAH for the
NMSSM are a precise mass-spectrum calculation using the full two-loop
RGEs of the NMSSM, with all flavor and CP effects,  for the running between $M_Z$, $M_{SUSY}$ and
  $M_{GUT}$.  The RGEs are
calculated using generic expressions given in
Refs.~\cite{Martin:1993zk,Yamada:1994id,Jack:1999zs,Jack:2000nm,Fonseca:2011vn,Goodsell:2012fm,Sperling:2013eva,Sperling:2013xqa}. \SPheno
calculates also the full one-loop corrections to all masses, including
the entire momentum dependence in the \DR scheme. At two-loop all
corrections in the gaugeless limit with vanishing external momenta are
included~\cite{Goodsell:2014bna,Goodsell:2015ira}, by making use of
generic expressions given in
Refs.~\cite{Martin:2003it,Martin:2003qz,Martin:2005eg}. This approach
makes it possible to include all two-loop corrections without making
use of any MSSM approximations as done by other codes, i.e.~also the
NMSSM-specific corrections of ${\cal O}(\alpha_\lambda
(\alpha_\lambda+\alpha_\kappa+\alpha_t))$ are fully included.
However, it is optionally also possible to make use of results from
the literature for the two-loop corrections in the MSSM and
NMSSM. These routines are usually much faster but have to be used
carefully for models beyond the MSSM.  Higgs sectors with CP violation
are so far supported only at the one-loop level.

Other features of the \SPheno NMSSM version are the calculation of all
important quark-flavor violating observables ($B$ and $K$ decays,
$\Delta M_{B_{d,s}}$/$\Delta M_K$, $\epsilon_K$) as well as
lepton-flavor violating observables using the {\tt FlavorKit}
functionality \cite{Porod:2014xia}. Also, $(g-2)_l$ and
electromagnetic dipole moments are predicted. Moreover, all two- and
three-body decays of SUSY particles, and two-body decays of the Higgs
scalars are calculated. The sparticle decays are purely tree-level,
while for the Higgs bosons the NLO-QCD corrections to decays in two
quarks, photons and gluons are included. Also decays in virtual vector
bosons are taken into account. Finally, \SPheno can make a prediction
for the electroweak fine-tuning according to measures proposed in
Refs.~\cite{Ellis:1986yg,Barbieri:1987fn} and writes all necessary
input files to test points with {\tt HiggsBounds}
\cite{Bechtle:2008jh,Bechtle:2011sb,Bechtle:2013wla} and {\tt
  HiggsSignals} \cite{Bechtle:2013xfa}.

The basic \SPheno version that is used together with \SARAH can be
downloaded from
  \begin{center}
  {\tt http://spheno.hepforge.org/} \\
 \end{center}
 A handy possibility to interface \SPheno versions created by \SARAH
 with Monte-Carlo tools like {\tt MadGraph}
 \cite{Alwall:2011uj,Alwall:2014hca} and {\tt WHIZARD}
 \cite{Kilian:2007gr,Moretti:2001zz} as well as {\tt MicrOmegas} is
 available via the {\tt BSM Toolbox} \cite{Staub:2011dp}.

\subsubsection{\NMSSMCalc}
\NMSSMCalc allows for the computation of the Higgs masses both in the
CP-conserving and CP-violating NMSSM, both at one-loop and at two-loop
level.  In contrast to the other codes, \NMSSMCalc makes use of mixed
\DR--OS renormalization conditions for
the computation of the Higgs masses. In the (s)top sector the user can
choose between OS and \DR renormalization. In addition, if the charged
Higgs boson mass $M_{H^{\pm}}$ is given as an input instead of
$A_{\lambda}$, $M_{H^{\pm}}$ enters the Higgs mass calculation as an
OS parameter. For comparison between the codes, however, 
the option of \DR renormalization of the (s)top sector and the option
with the input $A_{\lambda}$ are chosen. The Higgs mass calculation at one-loop
level is performed including the full momentum dependence and all
possible contributions \cite{Ender:2011qh,Graf:2012hh}. At the
two-loop level the ${\cal O}(\alpha_s \alpha_t)$ corrections are included
\cite{Muhlleitner:2014vsa}. They include the ${\cal O}(\alpha_s \alpha_t)$
part relating the vacuum expectation value to physical observables,
which is missing so far in the other spectrum generators. 

\NMSSMCalc can be downloaded from
\begin{center}
  {\tt http://www.itp.kit.edu/$\sim$maggie/NMSSMCALC/}
\end{center}
and comes together with an NMSSM extension of \HDECAY
\cite{Djouadi:1997yw,Djouadi:2006bz,Butterworth:2010ym} for the Higgs
boson decays.  The decays are calculated in the CP-conserving and
CP-violating NMSSM. The decay widths include the dominant higher-order
QCD corrections. Furthermore, the
decays of the neutral Higgs bosons into a bottom pair include the
higher order SUSY-QCD and the approximate SUSY-electroweak (EW)
corrections up to one-loop order. The decays into a strange quark pair
include the dominant resummed SUSY-QCD and the ones into $\tau$ pairs 
the dominant resummed SUSY-EW corrections. Accordingly the charged
Higgs boson decays into fermion pairs include the higher order SUSY
corrections.  All two-body decays into SUSY particles have been
implemented. For the CP-conserving case the decays into stop and
sbottom pairs come with the SUSY-QCD corrections. Finally,
all relevant off-shell decays into two massive gauge boson final
states, into gauge and Higgs boson final states, into Higgs pairs as
well as into heavy quark pairs are included. For more details and a
complete list of references see \cite{Baglio:2013iia}.

\subsubsection{\SoftSUSY}
\SoftSUSY is a widely used spectrum generator for the MSSM with and
without $R$-parity violation which has recently been extended to the
NMSSM.  The NMSSM spectrum generator is implemented to solve scenarios
with and without the $\mathbb{Z}_3$ symmetry.  The implementation of
 EWSB conditions allows
one to choose as output for the $\mathbb{Z}_3$-conserving
case either the set $\{ \kappa, v_s, m_s^2 \}$ or the
  set $\{ m_{H_u}^2, m_{H_d}^2, m_s^2 \}$, and in the
$\mathbb{Z}_3$-violating case there is an additional option of giving
out $\{ \mu, B_\mu, \chi_s \}$. For the high-scale boundary conditions
there are a number of pre-defined options including the
mSUGRA-inspired semi-constrained NMSSM and a general high-scale
boundary condition which allows one to set all soft
SUSY-breaking parameters independently.  The users may also
easily create their own boundary conditions.

For the NMSSM, \SoftSUSY extends the full three-family one- and
two-loop RGEs of the MSSM~\cite{Martin:1993zk,Barger:1993gh} 
using general two-loop expressions from the literature
\cite{Martin:1993zk,Yamada:1994id}, with the NMSSM specialization
given in Appendix D of the NMSSM manual~\cite{Allanach:2013kza}. These
expressions were also cross-checked against two-loop NMSSM RGEs in the
literature that use the 3rd-family
approximation~\cite{Ellwanger:2009dp}, and tested numerically against
the RGEs from \SARAH using an early development code from
\FlexibleSUSY. Finally, the RGEs for the vacuum expectation values are
implemented using Refs.~\cite{Sperling:2013eva,Sperling:2013xqa}.
The one-loop self-energies and tadpole corrections for the Higgs
bosons were extended to the NMSSM using the expressions in
Refs.~\cite{Degrassi:2009yq,Staub:2010ty}. The NMSSM extension also
includes the two-loop corrections at order ${\cal O}(\alpha_s(\alpha_t
+ \alpha_b))$ from Ref.~\cite{Degrassi:2009yq} and in addition uses
the MSSM results from
Refs.~\cite{Brignole:2001jy,Dedes:2002dy,Dedes:2003km,Allanach:2004rh}.

Higgs and sparticle decays may be obtained by interfacing SoftSUSY
with \NMHDECAY \cite{Ellwanger:2004xm,Ellwanger:2005dv} and \NMSDECAY
\cite{Das:2011dg}, which is based upon Ref.~\cite{Muhlleitner:2003vg}.
These are both distributed as part of the \NMSSMTools package, and
\SoftSUSY provides a script to do the interface with this package
automatically.

The homepage of \SoftSUSY is  
\begin{center}
  {\tt http://softsusy.hepforge.org/}
 \end{center}

\subsubsection{\FlexibleSUSY}

\FlexibleSUSY is a Mathematica and C++ package, which creates spectrum
generators for user-defined (SUSY or non-SUSY) models.  It makes use
of \SARAH to obtain model-dependent details, i.e.~couplings,
RGEs, self-energies and tadpole
corrections.  The boundary conditions on the model parameters at
different scales, as well as spectrum-generator specific configuration
details are specified in a \FlexibleSUSY model file.
\FlexibleSUSY can therefore generate spectrum generators for the
different MSSM singlet extensions provided by \SARAH, for example the
NMSSM with and without a $\mathbb{Z}_3$ symmetry.  The generated spectrum
generators consist of C++ classes, which can be easily adapted
allowing one to use the routines in a custom-made program for solving
non-standard problems.  For example one may build a tower of effective
field theories, which could be useful for NMSSM scenarios with very
heavy superparticles.
Currently, \FlexibleSUSY fully supports only
real parameters. Complex parameters have been developed very recently
in version {\tt 1.1.0}, but they are still undocumented and
undergoing testing. The automatic implementation of alternative
boundary-value-problem solvers and automatic generation of 
towers of effective field theories are
currently under development. Additionally, an extension to calculate
decays, \FlexibleDecay, for both Higgs and superparticles is currently
being developed.

Due to the use of \SARAH, \FlexibleSUSY employs the full two-loop RGEs
and one-loop corrections to the pole mass spectrum.  At the two-loop
level it makes use of Higgs mass corrections available in the
literature.  In the case of the NMSSM \FlexibleSUSY uses the
$\alpha_s(\alpha_t + \alpha_b)$ corrections given in
Ref.~\cite{Degrassi:2009yq}.  In addition, the MSSM two-loop
corrections based on
Refs.~\cite{Brignole:2001jy,Dedes:2002dy,Dedes:2003km,Allanach:2004rh}
are used, though the user may choose to switch any of these two-loop
corrections off using the SLHA file.

The homepage of \FlexibleSUSY is
 \begin{center}
  {\tt http://flexiblesusy.hepforge.org/}
 \end{center}

\bigskip

\subsection{Used codes and options}
For the comparison in the following, these versions of the different
codes were used:
\begin{itemize}
\item \NMSSMTools\ \tt 4.7.0 
 \item \SPheno\ {\tt 3.3.6} together with \SARAH\ {\tt 4.5.7}
 \item \NMSSMCalc\ {\tt 1.0.3} 
 \item \SoftSUSY\ {\tt 3.6.0} 
 \item \FlexibleSUSY\ {\tt 1.1.0} together with
   \SARAH\ {\tt 4.5.7}
\end{itemize}
We always make use of the most precise calculation available in each code, i.e. all available two-loop corrections turned on  
and in \NMSSMTools we only compare Higgs masses calculated using option $1$, described in section \ref{sec:NMSSMTools}, and do not consider any of the approximations from alternative options.
We also concentrate on calculations done in the \DR scheme. A discussion of the effects using an OS scheme
is beyond the scope of this comparison and will be given elsewhere including also 
results from the upcoming NMSSM version of {\tt FeynHiggs} \cite{Heinemeyer:1998yj,Heinemeyer:1998np,Degrassi:2002fi,Frank:2006yh,Hahn:2013ria}.

\subsection{Predictions for the CP-even and CP-odd Higgs masses}
\label{sec:OriginalMasses}
We discuss here the Higgs mass predictions of the various codes for
six different test points (TP) with the following features:
\begin{itemize}
 \item MSSM-like point  (TP1)
 \item MSSM-like point with large stop splitting (TP2)
 \item Point with light singlet and $\lambda$ close to the perturbativity limit (TP3)
 \item Point with heavy singlet and $\lambda$ close to the perturbativity limit (TP4)
 \item Point with slightly lighter singlet. Additional matter is needed for perturbativity; inspired by NMP9 \cite{King:2012is} (TP5)
 \item Point with huge $\lambda$ (TP6)
\end{itemize}

In table~\ref{tab:input} we list the input parameters that are
different for the six benchmark points. As is usual in
    NMSSM studies, for the last four points we choose smallish values
    of $\tan\beta$, to avoid suppressing the NMSSM-specific
    contributions to the SM-like Higgs mass. For very large values of
$\tan\beta$ additional differences
  between the codes are expected, since not all of them
  include the two-loop corrections
  involving the bottom and $\tau$ Yukawa coupling.

All values in this table apart from the one for
$\tan\beta$ are defined at the scale $Q$. The shown value for
$\tan\beta$ refers to the scale $M_Z$~\footnote{Since \NMSSMCalc and
  \NMSSMTools expect $\tan\beta$ to be defined at 
  $Q$ as well, the input values have to be adjusted
  accordingly.}. The remaining soft SUSY-breaking parameters,
  common to all points, are
%
\begin{eqnarray}
& M_{\tilde f} = 1500~\gev~, \,  A_\tau = 0~. &
\end{eqnarray}
In addition, if possible via the input file, the codes were
forced to calculate the masses directly at the
input scale $Q$. This is possible for all tools except
\NMSSMTools. More details about the general handling of the scales in
the different codes will be given in section~\ref{sec:scales}.
These points are chosen in a way that differences between the 
predicted Higgs masses are visible. However, they are not the result of
an exhaustive scan to find points with the largest possible
differences. Thus, we expect that these points are representative
of a non-negligible fraction of the parameter space
of the NMSSM.
\begin{table}[t] 
\begin{tabular}{|c|c|cccccc|ccccccc|}
\hline
& $Q$ & $\tan\beta$ & $\lambda$ & $\kappa$ & $A_\lambda$ & $A_\kappa$ & $\mu_{eff}$ & $M_1$ & $M_2$ & $M_3$ & $A_t$ & $A_b$ & $m_{\tilde t_L}$ & $m_{\tilde t_R}$ \\ 
\hline
TP1 & {1500}&{10}&{0.1}&{0.1}&{-10}&{-10}&{900}&{500}&{1000}&{3000}&{3000}&{0}&{1500}&{1500}\\ 
TP2 & {1500}&{10}&{0.05}&{0.1}&{-200}&{-200}&{1500}&{1000}&{2000}&{2500}&{-2900}&{0}&{2500}&{500}\\ 
TP3 & {1000}&{3}&{0.67}&{0.1}&{650}&{-10}&{200}&{200}&{400}&{2000}&{1000}&{1000}&{1000}&{1000}\\ 
TP4 & {750}&{2}&{0.67}&{0.2}&{405}&{0}&{200}&{120}&{200}&{1500}&{1000}&{1000}&{750}&{750}\\ 
TP5 & {1500}&{3}&{0.67}&{0.2}&{570}&{-25}&{200}&{135}&{200}&{1400}&{0}&{0}&{1500}&{1500}\\ 
TP6 & {1500}&{3}&{1.6}&{1.61}&{375}&{-1605}&{614}&{200}&{400}&{2000}&{0}&{0}&{1500}&{1500}\\ 
\hline
\end{tabular}
\caption{Input parameters for the test points. All dimensionful parameters in units of \gev.}
\label{tab:input}
\end{table}

TP1 and TP2 have small couplings between the Higgs doublets and the
singlet. Thus, a mass of
about 125~GeV for the SM-like state can only be obtained via
large (s)top corrections as in the MSSM. In contrast, TP3-TP6 come
with a large $\lambda$ coupling and a small value for
$\tan\beta$. This already increases the mass of the SM-like scalar at tree level. However,
additional loop corrections beyond the MSSM are also expected to
become more important. The difference between these four points is
that TP3 and TP4 have values for $\lambda$ and $\kappa$ that
give a consistent model without any Landau pole up to the
GUT scale of $10^{16}$~GeV. For TP5 $\lambda$ becomes non-perturbative
below the GUT scale unless additional matter is assumed to change the
running. TP6 has a huge $\lambda$ which always leads to a
cut-off. Scenarios with these large $\lambda$ couplings can be
motivated by naturalness considerations~\cite{Hall:2011aa}.
 
\begin{table}[hbt]
\centering
\begin{tabular}{|l|cccccc|}
\hline
& TP1 & TP2 & TP3 & TP4 & TP5 & TP6 \\
\hline
 & \multicolumn{6}{|c|}{$h_1$}\\
\hline 
{\tt FlexibleSUSY }  & {\bf 123.55} & {\bf 122.84}       & {\it 91.11}      & {\bf 127.62}       & {\it 120.86}     &  {\bf 126.46}                                                                                  \\ 
{\tt NMSSMCALC }     & {\bf 120.34}  & {\bf 118.57}       & {\it 90.88}      & {\bf 126.37}        & {\it 120.32}     &  {\bf 123.45}                                                                                  \\ 
{\tt NMSSMTOOLS }    & {\bf 123.52}   & {\bf 121.83}       & {\it 90.78}      & {\bf 127.30}       & {\it 119.31}     &  {\bf 126.63}                                                                                       \\ 
{\tt SOFTSUSY }      & {\bf 123.84}   & {\bf 123.08}       & {\it 90.99}      & {\bf 127.52}        & {\it 120.81}     & {\bf 126.67}                                                                                       \\ 
{\tt SPHENO }        & {\bf 124.84}   & {\bf 124.74}       & {\it 89.54}      & {\bf 126.62}        & {\it 119.11}     & {\bf 131.29}                                                                                       \\ 
\hline
 & \multicolumn{6}{|c|}{$h_2$}\\
\hline 
{\tt FlexibleSUSY }  & {\it 1797.46}  & {\it 5951.36}      & {\bf 126.58}     & {\it 143.11}       & {\bf 125.08}      & {\it 700.80}                                                                                      \\ 
{\tt NMSSMCALC }     & {\it 1797.46}  & {\it 5951.36}      & {\bf 124.86}     & {\it 142.59}     & {\bf 123.14}      & {\it 701.02}                                                                                   \\ 
{\tt NMSSMTOOLS }    & {\it 1797.46}  & {\it 5951.36}      & {\bf 127.28}     & {\it 144.07}       & {\bf 126.95}      & {\it 700.46}                                                                                         \\ 
{\tt SOFTSUSY }      & {\it 1797.46}  & {\it 5951.36}      & {\bf 126.59}     & {\it 143.02}       & {\bf 125.12}      & {\it 701.01}                                                                                         \\ 
{\tt SPHENO }        & {\it 1798.01}  & {\it 5951.35}      & {\bf 126.77}     & {\it 144.04}       & {\bf 125.61}      & {\it 689.30}                                                                                         \\ 
\hline
 & \multicolumn{6}{|c|}{$h_3$}\\
\hline 
{\tt FlexibleSUSY }  & {2758.96}  & {6372.08}       & {652.95}     & {467.80}       &{627.28}     & {1369.53}                                                                                     \\ 
{\tt NMSSMCALC }     & {2756.37}   & {6371.31}      & {652.48}    & {467.42}       &{627.00}     & {1368.68}                                                                                 \\ 
{\tt NMSSMTOOLS }    & {2758.51}  & {6345.72}      & {651.03}     & {466.38}        &{623.79}     & {1368.90}                                                                                         \\ 
{\tt SOFTSUSY }      & {2758.41}  & {6370.3}      & {652.78}     & {467.73}        & {627.14}     & {1369.19}                                                                                         \\ 
{\tt SPHENO }        & {2757.11}  & {6366.88}      & {651.21}     & {467.5}        & {624.02}     & {1363.02}                                                                                         \\ 
\hline
\hline
\end{tabular}
\caption{Masses for the CP-even scalars (in GeV) for TP1--TP6 when
  using the spectrum generators ``out-of-the-box''. The values
  correspond to the two-loop results obtained by the different
  tools. For \NMSSMCalc the value using the \DR\ scheme is shown and
  for \NMSSMTools the value using the most precise calculation is
  given.  The masses for the SM-like scalar
  are written in bold fonts, those for the singlet-like scalar
  in italics.}
\label{tab:original}
\end{table}

\begin{table}[htb]
\centering
\begin{tabular}{|l|ccc|ccc|ccc|}
\hline
 & \multicolumn{3}{|c|}{TP1}   &  \multicolumn{3}{|c|}{TP2}  & \multicolumn{3}{|c|}{TP3} \\
 & $|Z_{i1}|$ & $|Z_{i2}|$ &  $|Z_{i3}|$      &  $|Z_{i1}|$ & $|Z_{i2}|$ &  $|Z_{i3}|$       &  $|Z_{i1}|$ & $|Z_{i2}|$ &  $|Z_{i3}|$  \\
\hline
{\tt FlexibleSUSY } &{0.1039} & {0.9946} & {0.0076}&{0.1034} & {0.9946} & {0.0004} &{0.2172} & {0.1888} & {0.9577} \\ 
{\tt NMSSMCALC }  &{0.1039} & {0.9946} & {0.0076}  &{0.1034} & {0.9946} & {0.0004} &{0.2177} & {0.1923} & {0.9569} \\ 
{\tt NMSSMTOOLS }  & {0.1039} & {0.9946} & {0.0076}& {0.1038} & {0.9946} & {0.0004}& {0.2229} & {0.2115} & {0.9511} \\ 
{\tt SOFTSUSY }  &{0.1039} & {0.9946} & {0.0076}   &{0.1034} & {0.9946} & {0.0004} &{0.2164} & {0.1883} & {0.958} \\ 
{\tt SPHENO }   &{0.104} & {0.9945} & {0.0074}     &{0.1035} & {0.9946} & {0.0004} &{0.2269} & {0.2265} & {0.9472} \\ 
\hline
{\tt FlexibleSUSY } &{0.0075} & {0.0069} & {0.9999}&{0.0096} & {0.0006} & {1.}     &{0.2814} & {0.9274} & {0.2466} \\ 
{\tt NMSSMCALC }  &{0.0075} & {0.0068} & {0.9999}  &{0.0095} & {0.0006} & {1.}     &{0.2811} & {0.9265} & {0.2502} \\ 
{\tt NMSSMTOOLS }  & {0.0076} & {0.0069} & {0.9999}& {0.0101} & {0.0006} & {0.9999}& {0.2782} & {0.9216} & {0.2708} \\ 
{\tt SOFTSUSY }  &{0.0075} & {0.0068} & {0.9999}   &{0.0096} & {0.0006} & {1.}      &{0.282} & {0.9273} & {0.246} \\ 
{\tt SPHENO }   &{0.0075} & {0.0067} & {0.9999}    &{0.0096} & {0.0006} & {1.}     &{0.2725} & {0.919} & {0.2851} \\ 
\hline
{\tt FlexibleSUSY } &{0.9946} & {0.1039} & {0.0068}&{0.9946} & {0.1034} & {0.0096} &{0.9347} & {0.3231} & {0.1483} \\ 
{\tt NMSSMCALC }  &{0.9946} & {0.104} & {0.0068}   &{0.9946} & {0.1034} & {0.0095} &{0.9347} & {0.3234} & {0.1476} \\ 
{\tt NMSSMTOOLS }  & {0.9946} & {0.104} & {0.0068} & {0.9945} & {0.1038} & {0.0102}& {0.9338} & {0.3255} & {0.1484} \\ 
{\tt SOFTSUSY }  &{0.9946} & {0.104} & {0.0068}    &{0.9946} & {0.1034} & {0.0096} &{0.9347} & {0.3234} & {0.1476} \\ 
{\tt SPHENO }   &{0.9945} & {0.104} & {0.0068}     &{0.9946} & {0.1035} & {0.0097} &{0.935} & {0.3228} & {0.1468} \\ 
\hline
\end{tabular}
\caption{
Absolute values of the entries of the rotation matrix
  $Z_H$ corresponding to the
    SM-like scalar for TP1--TP3 when using the spectrum generators
  ``out-of-the-box''.  The values
  correspond to the two-loop results obtained by the different
  tools. For \NMSSMCalc the value using the \DR\ scheme is shown and
  for \NMSSMTools the value using the most precise calculation is
  given. }
\label{tab:originalZH13}
\end{table}

\begin{table}[htb]
\centering
\begin{tabular}{|l|ccc|ccc|ccc|}
\hline
 & \multicolumn{3}{|c|}{TP4}   &  \multicolumn{3}{|c|}{TP5}  & \multicolumn{3}{|c|}{TP6} \\
 & $|Z_{i1}|$ & $|Z_{i2}|$ &  $|Z_{i3}|$      &  $|Z_{i1}|$ & $|Z_{i2}|$ &  $|Z_{i3}|$       &  $|Z_{i1}|$ & $|Z_{i2}|$ &  $|Z_{i3}|$  \\
\hline
{\tt FlexibleSUSY } &{0.4866} & {0.7908} & {0.3712}  &{0.2867} & {0.4021} & {0.8695} &{0.2741} & {0.9417} & {0.195} \\ 
{\tt NMSSMCALC }  &{0.4869} & {0.7849} & {0.3832}    &{0.2967} & {0.4433} & {0.8459} &{0.2746} & {0.9416} & {0.1947} \\ 
{\tt NMSSMTOOLS }  & {0.4855} & {0.7495} & {0.4501}  & {0.3072} & {0.4711} & {0.8269} & {0.2743} & {0.9416} & {0.1953} \\ 
{\tt SOFTSUSY }  &{0.4867} & {0.7902} & {0.3725}     &{0.2863} & {0.402}  & {0.8697} &{0.2747} & {0.9417} & {0.1941} \\ 
{\tt SPHENO }   &{0.4835} & {0.7646} & {0.4262}      &{0.3033} & {0.4703} & {0.8288} &{0.2712} & {0.9407} & {0.2039} \\ 
\hline                                                                               
{\tt FlexibleSUSY } &{0.0370} & {0.4059} & {0.9132}  &{0.2189} & {0.8561} & {0.4681} &{0.2927} & {0.1115} & {0.9497} \\ 
{\tt NMSSMCALC }  &{0.0447} & {0.4158} & {0.9084}    &{0.2055} & {0.8354} & {0.5099} &{0.2919} & {0.1113} & {0.9500} \\ 
{\tt NMSSMTOOLS }  & {0.0834} & {0.4727} & {0.8772}  &{0.1974} & {0.8184} & {0.5396} & {0.2937} & {0.1114} & {0.9494} \\ 
{\tt SOFTSUSY }  &{0.0385} & {0.4066} & {0.9128}     &{0.2195} & {0.8561} & {0.4679} &{0.2918} & {0.1107} & {0.9501} \\ 
{\tt SPHENO }   &{0.0682} & {0.4525} & {0.8892}      &{0.1958} & {0.8204} & {0.5372} &{0.2945} & {0.1206} & {0.948} \\ 
\hline                                                                               
{\tt FlexibleSUSY } &{0.8728} & {0.4581} & {0.1683}  &{0.9327} & {0.3245} & {0.1575} &{0.9161} & {0.3174} & {0.2451} \\ 
{\tt NMSSMCALC }  &{0.8723} & {0.4594} & {0.1674}    &{0.9326} & {0.3251} & {0.1568} &{0.9162} & {0.3177} & {0.2443} \\ 
{\tt NMSSMTOOLS }  & {0.8703} & {0.4634} & {0.1670}  &{0.9309} & {0.3291} & {0.1584} & {0.9157} & {0.3177} & {0.2460} \\ 
{\tt SOFTSUSY }  &{0.8727} & {0.4587} & {0.1675}     &{0.9327} & {0.3249} & {0.1568} &{0.9162} & {0.3176} & {0.2444} \\ 
{\tt SPHENO }   &{0.8727} & {0.4589} & {0.1667}      &{0.9326} & {0.3252} & {0.1567} &{0.9164} & {0.3172} & {0.2443} \\ 
\hline
\end{tabular}
\caption{The same as Tab.~\ref{tab:originalZH46} for TP4--TP6. }
\label{tab:originalZH46}
\end{table}

The scalar masses calculated for these six points
by the five spectrum generators as they come
  ``out-of-the-box'' are listed in table~\ref{tab:original},
and the absolute values of the entries of the rotation matrix
  $Z_H$ that correspond to the SM-like scalar are listed in
tables~\ref{tab:originalZH13}-\ref{tab:originalZH46}. For TP1, TP2 and TP6 the
singlet-like scalar is very heavy and decouples, for TP4 the
singlet-like scalar is just slightly heavier than the SM-like
scalar, while for TP3 and TP5 it is the
lightest scalar. We find that, even for the MSSM-like points, the
differences in the mass of the
  SM-like scalar can exceed 5~GeV. For the points with very large
$\lambda$ the difference can be even larger. The masses for the other
two scalars usually agree quite well between the
codes. Only \SPheno gives masses of the singlet-like
states, for TP5 and TP6, which differ by 1--2\% compared to all
other tools.

\section{Breaking down the differences}
\label{sec:differences}
\subsection{Main differences}
The differences we have observed are larger than one usually finds for
the MSSM when comparing the widely used tools that perform a \DR\ calculation
\cite{Allanach:2004rh,Allanach:2003jw}. We pinned down four sources
which will be discussed in the following. The numerical impact of these 
differences is discussed in sec.~\ref{sec:size}.
Further differences, such as
e.g.~whether first and second generation fermions are included into
the one-loop computation (\NMSSMCalc, \SPheno, \FlexibleSUSY) or not
(\NMSSMTools, \SoftSUSY), are numerically tiny.

\subsubsection{Renormalization scheme}
All codes but \NMSSMCalc perform a pure \DR
renormalization. \NMSSMCalc uses instead mixed \DR-OS
renormalization conditions at one-loop level. At two-loop level it
offers an option to renormalize the (s)top sector not only OS
but also \DR.

\subsubsection{Calculation of the running  parameters}
\label{subsec:calculation-of-drbar-parameters}
A very crucial point is how the \DR parameters for the gauge and
Yukawa couplings at the SUSY scale are calculated, since these
couplings affect all loop calculations.

\paragraph{Top Yukawa coupling}
The most important parameter entering the loop calculations in the
Higgs sector is the top Yukawa coupling. To calculate the
value of the \DR-renormalized Yukawa coupling
at the SUSY scale, there are two different
approaches implemented in the codes:

\begin{enumerate}
\item {\bf SUSY thresholds at $M_Z$:~} \FlexibleSUSY, \SoftSUSY and
  \SPheno adapt to the NMSSM the approach presented in
  Ref.~\cite{Pierce:1996zz} to calculate the running 
  parameters of the MSSM Lagrangian at the SUSY scale.
  In this setup all one-loop SUSY and non-SUSY
  thresholds are calculated at the weak scale, identified with
    the $Z$-boson mass $M_Z$, and also the known two-loop QCD
  corrections of \cite{Avdeev:1997sz,Bednyakov:2002sf} are
  included. The evolution of the parameters
  from $M_Z$ to the SUSY scale $M_{SUSY}$ is done using two-loop SUSY
  RGEs. A detailed description of the calculation is given in
  Ref.~\cite{Athron:2014yba} for \FlexibleSUSY,
  Ref.~\cite{Allanach:2013kza} for \SoftSUSY and in
  Ref.~\cite{Staub:2015kfa} for \SPheno. 
  All three codes use in principle the same approach:

 the \DR mass of the top quark
   is computed as
   \begin{equation}
 \label{mfermion}
     m_t^{\DR} = m_t^{pole} + \hat{m}_t \,\Sigma(\hat{m}_t^2) \;, 
   \end{equation}
  where $\Sigma(p^2)$ is the (dimensionless) self-energy 
  for momentum $p$.\footnote{More precisely, \SPheno calculates
   the chiral and scalar self-energy contributions $\Sigma_L$,
   $\Sigma_R$, $\Sigma_S$ to the $3 \times 3$ mass matrix, but this
   only leads to differences if flavor violation is included. }
 For other fermions, the relation becomes more complicated
   because of the $\tan\beta$ resummation which is performed by the
   codes.  The \DR Yukawa couplings are then computed at $M_Z$ by
 dividing the \DR fermion masses by the appropriate VEVs. There is
 however a difference between the codes in the definition of the mass
 $\hat m_{{t}}$ entering the correction in eq.~(\ref{mfermion}):
 \SoftSUSY and \FlexibleSUSY use $ \hat{m}_{{t}} =
 m_{{t}}^{pole}$, while \SPheno uses $ \hat{m}_{{t}} =
 m_{{t}}^{\DR}$.  The effect
 of the different definition of $\hat m_{{t}}$ in the one-loop
 part of correction is partially compensated for by an appropriate
 shift in the two-loop QCD part of the correction.

  \item {\bf SUSY thresholds at $M_{SUSY}$:~} In this scheme, the
    pole top mass is converted to the \MS\ scheme at
    the scale $m_t$, and the RGEs of the SM
    are used for the running to $M_{SUSY}$. At $M_{SUSY}$ the
    conversion of the 
    \MS\ top mass of the SM to the corresponding \DR Yukawa
      coupling of the NMSSM takes place including the SUSY
    thresholds. This is the approach followed in \NMSSMTools and
    \NMSSMCalc. However, there remain some differences in the concrete
    realization. \NMSSMCalc uses the NLO-QCD contributions to the top-mass RGE for the running
    up to $M_{SUSY}$, and includes only the SQCD threshold
    corrections at $M_{SUSY}$.  \NMSSMTools resums also large
    logarithms $\sim Y_t^2\log(M_{SUSY}/m_t)$ and $\sim
    Y_t^2\log(M_{SUSY}/M_A)$ (the latter in case the scale
    $M_A$ of the heavy MSSM-like scalars differs from the scale
    $M_{SUSY}$ of the stop/sbottom) and it includes also
      ${\cal O}(\alpha_t)$ threshold corrections from stop-Higgsino
      loops at $M_{SUSY}$.
  \end{enumerate}

\paragraph{Strong interaction coupling}
The strong interaction coupling at the
  scale $M_{SUSY}$, important for the two-loop corrections to
  the Higgs masses, is derived as follows:
\begin{itemize}
\item \FlexibleSUSY, \SoftSUSY, \SPheno calculate $\alpha_s^{\DR}(M_Z)$ via 
  \begin{align}
    \alpha_s^{\DR}(M_Z) &= \frac{\alpha_s^{(5),\overline{\text{MS}}}(M_Z)}{1 - \Delta\alpha_s(M_Z)},  \\
    \Delta\alpha_s(M_Z) =& \frac{\alpha_s}{2\pi} \left( \frac{1}{2} -
      2 \log{\frac{m_{\tilde g}}{M_Z}} - \frac{2}{3}
      \log{\frac{m_t}{M_Z}} - \frac{1}{6} \sum_{i=1}^6 \left(
        \log{\frac{m_{\tilde u_i}}{M_Z}} + \log{\frac{m_{\tilde
              d_i}}{M_Z}} \right) \right),
  \end{align}
  where $\alpha_s^{(5),\overline{\text{MS}}}$ denotes the strong
  coupling constant of the SM with $N_f=5$ active flavors.  Finally,
  $\alpha_s^{\DR}(M_Z)$ is run to $M_{SUSY}$ using the two-loop 
  RGEs of the NMSSM.

\item \NMSSMCalc: $\alpha_s^{\DR}(M_{SUSY})$ is computed by running
  $\alpha_s^{(5),\overline{\text{MS}}}(M_Z)$ up to the top pole mass
  $m_t$ using the QCD part of the two-loop RGEs of the SM with $N_f=5$
  active flavors.  Afterwards, the so-obtained
  $\alpha_s^{\overline{\text{MS}}}(m_t)$ is run to $M_{SUSY}$ using
  the corresponding RGEs with $N_f=6$ active flavors.  With this
  procedure the one-loop top threshold correction for
  $\alpha_s^{\overline{\text{MS}}}$ is taken into account implicitly.
  Finally, $\alpha_s^{\overline{\text{MS}}}(M_{SUSY})$ is converted to
  \DR by
  \begin{equation}
    \alpha_s^{\DR}(M_{SUSY})=\frac{\alpha_s^{\overline{\text{MS}}}(M_{SUSY})}{1-\frac{\alpha_s^{\overline{\text{MS}}}(M_{SUSY})}{4 \pi}}\,.
  \end{equation}

\item 
 \NMSSMTools: $\alpha_s^{\overline{\text{MS}}}(M_{SUSY})$ is
  computed with a procedure analogous to the one of \NMSSMCalc, but it
  is not converted to the \DR\ scheme. This difference amounts to a
  three-loop (i.e., higher-order) effect in the calculation of the
  Higgs masses.

\end{itemize}

\paragraph{Electroweak interactions} 
 The EW gauge sector of the NMSSM is determined by three
  fundamental parameters, i.e.~the gauge couplings\footnote{ For the
    $U(1)$ coupling $g_1$ we adopt the SM normalization, i.e.~$g_1 =
    g^\prime$.} $g_1$ and $g_2$ and the combination of Higgs VEVs $v
  = (v_u^2+v_d^2)^{1/2}$. For a consistent determination of these
  three parameters, a choice of three physical inputs is required. The
  procedures adopted by the five codes differ in the choices of physical
  inputs, as well as, in the case of \NMSSMCalc, the choice of
  renormalization scheme for the fundamental parameters.

\begin{enumerate}
\item \FlexibleSUSY, \SoftSUSY and \SPheno adapt again the approach of
  Ref.~\cite{Pierce:1996zz} to calculate $g_1$ and $g_2$. The
    three physical inputs are chosen as the $Z$ mass, the Fermi
    constant $G_F$ and the electromagnetic coupling of the SM at the
    scale $M_Z$ in the 5-flavor scheme,
    $\alpha_{em}^{(5),\overline{\text{MS}}}(M_Z)$.  The threshold
  corrections to $\alpha_{em}$ from SUSY states, the charged Higgs and
  the top are calculated at the scale $M_Z$ as
\begin{align}
  \alpha_{em}^{\DR}(M_Z) &= \frac{\alpha_{em}^{(5),\overline{\text{MS}}}(M_Z)}{1 - \Delta\alpha(M_Z)} ,\\
    \Delta\alpha(M_Z) =& \frac{\alpha}{2\pi} \Big(\frac{1}{3}-
      \frac{16}{9}  \log{\frac{m_{t}}{M_Z}} - \frac{4}{9} \sum_{i=1}^6
      \log{\frac{m_{\tilde u_i}}{M_Z}} - \frac{1}{9} \sum_{i=1}^6
      \log{\frac{m_{\tilde d_i}}{M_Z}}  \nonumber \\
      & \hspace{1cm}  -\frac{4}{3} \sum_{i=1}^2
      \log{\frac{m_{\tilde \chi^+_i}}{M_Z}} - 
      \frac{1}{3} \sum_{i=1}^6 \log \frac{m_{\tilde{e}_i}}{M_Z}
     -      \frac{1}{3}       \log{\frac{m_{H^+}}{M_Z}} \Big) .
\end{align}
The Weinberg angle $\sin^{\DR}\Theta_W$ at the scale $M_Z$ is
  obtained iteratively from the above-computed $ \alpha_{em}^{\DR}(M_Z)$,
    plus $G_F$ and $M_Z$, via the relation
\begin{align}
  \left(\sin^{\DR}\Theta_W \cos^{\DR}\Theta_W\right)^2 =
  \frac{\pi\,\alpha_{em}^{\DR}(M_Z)}{\sqrt{2} M_Z^2 G_F (1-\delta_r)} ,
\end{align}
where
\begin{align}
  \delta_r &= \hat\rho \frac{\Pi_{WW}^T(0)}{M_W^2} -
  \frac{\re\Pi_{ZZ}^T(M_Z^2)}{M_Z^2} + \delta_{\rm VB} + \delta_r^{(2)} , \\
  \hat\rho &= \frac{1}{1-\Delta\hat\rho} ,\qquad\qquad \Delta\hat\rho
  = \re\Biggl[ \frac{\Pi_{ZZ}^T(M_Z^2)}{\hat\rho\,M_Z^2} -
  \frac{\Pi_{WW}^T(M_W^2)}{M_W^2}\Biggr] + \Delta\hat\rho^{(2)}~,
\end{align}
$\Pi^T_{VV}(p^2)~~(V = Z,W)$ are the \DR-renormalized transverse parts
of the self-energies of the vector bosons, computed at the
renormalization scale $Q=M_Z$, and $\delta_r^{(2)}$ and
$\Delta\hat\rho^{(2)}$ are two-loop corrections given in
\cite{Fanchiotti:1992tu,Pierce:1996zz}.  The one-loop vertex and box
corrections $\delta_{\rm VB}$ implemented into \FlexibleSUSY and
\SoftSUSY are 
  from refs.~\cite{Degrassi:1990tu, Grifols:1984xs,
    Chankowski:1993eu}, generalized to
  give the complete one-loop result for the NMSSM, while the ones
used in \SPheno are automatically generated with \SARAH for the NMSSM.
We remark that, in this approach, the physical value of the
$W$ mass is an output of the calculation. At a generic scale $Q$, the
relation between the pole mass $M_V$ of a gauge boson $V$ and its
corresponding $\DR$ mass reads
\begin{align}
\label{gaugebosons}
M^{\DR}_V(Q) &~=~ \sqrt{M_V^2 ~+ ~\Pi^T_{VV}(M_V^2)}~,
\end{align}
where the self-energy is also computed at the scale $Q$.

In \FlexibleSUSY $\sin^{\DR}\Theta_W$ can be alternatively
calculated from the \DR $W$ and $Z$ masses, via
\begin{align}
  \sin^{\DR}\Theta_W(M_Z) = \sqrt{1
    - \left(\frac{M_W^{\DR}(M_Z)}{M_Z^{\DR}(M_Z)}\right)^2} .
  \label{eq:flexiblesusy-weinberg-angle-1.0.4}
\end{align}%
In this case the three physical inputs used to calculate $g_1$ and
$g_2$ are $M_Z$, $M_W$ and
$\alpha_{em}^{(5),\overline{\text{MS}}}(M_Z)$.  This approach is the
original one implemented in \FlexibleSUSY 1.0.0 and when we refer to
``out-of-the-box'' results for \FlexibleSUSY it will be using this
option.

The so-obtained values for $\alpha^{\DR}_{em}$ and $\sin^{\DR}
\Theta_W$ are used to get $g_1^{\DR}(M_Z)$ and
$g_2^{\DR}(M_Z)$, which are then evolved to the SUSY scale using
two-loop  RGEs of the NMSSM.

\item \NMSSMTools\ uses $M_Z$, $M_W$ and $G_F$ as physical
  inputs for the EW sector. It calculates the 
  \DR values of $g_1$ and $g_2$ at the scale $M_{SUSY}$
  by making use of the relations between the fundamental EW
    parameters and the \DR masses of the gauge bosons, the latter
    computed at the scale $Q=M_{SUSY}$ as in eq.~(\ref{gaugebosons}):
\begin{align}
\label{gauge1}
\sqrt{g_1^{\DR}(M_{SUSY})^2 ~+~ g_2^{\DR}(M_{SUSY})^2} ~~=~& 2 \,{M_Z^{\DR}(M_{SUSY})}
\,/\,{v^{\DR}(M_{SUSY})}~, \\[2mm]
g_2^{\DR}(M_{SUSY}) ~=~& 2\, {M_W^{\DR}(M_{SUSY})}\,/\,{v^{\DR}(M_{SUSY})}~,
\end{align}
with $v^{\DR}(M_{SUSY})$ calculated via
eq.~(\ref{eq:vNMSSMTOOLS})\footnote{All expressions assume $v \simeq
  246$~GeV.} below. 

\item \NMSSMCalc uses $M_Z$, $M_W$ and the  electromagnetic gauge coupling $e$ as physical inputs
  for the EW sector. Moreover, it does not compute \DR values for the
  EW gauge couplings, but adopts
the OS conditions
\begin{equation}
 g_1 ~=~ e \,\frac{M_Z}{M_W} \,, \hspace{1cm}
 g_2~=~\frac{e}{\sqrt{1-M_W^2/M_Z^2}} \;,
\end{equation}
where $M_W$, $M_Z$ and $e$ are the measured values. 

\end{enumerate}

\paragraph{Electroweak VEV}
The last SM parameter  that enters the calculation
of the Higgs masses is $v$, which
gets translated into the two Higgs VEVs $v_d$ and $v_u$ using the
input value of $\tan\beta$. The parameter $v$ at the
  scale $M_{SUSY}$ is obtained as follows in the different codes:

\begin{enumerate}
 \item \FlexibleSUSY and \SoftSUSY calculate the $\DR$ value at $M_Z$ as
 \begin{align}
   v^{\DR}(M_Z) &= \frac{2 \,M_Z^{\DR}(M_Z)}
{{\sqrt{g_1^{\DR}(M_Z)^2 ~+~ g_2^{\DR}(M_Z)^2}}}~.
 \end{align}

 $v^{\DR}(M_{SUSY})$ is then obtained using two-loop RGEs
with the $\xi$-dependent terms calculated in the Feynman gauge
\cite{Sperling:2013eva,Sperling:2013xqa}.

 \item \SPheno calculates $v^{\DR}$
   directly at $M_{SUSY}$ as
 \begin{equation}
{   v^{\DR}(M_{SUSY}) = \frac{2\, M_Z^{\DR}(M_{SUSY})}
{\sqrt{g_1^{\DR}(M_{SUSY})^2 ~+~ g_2^{\DR}(M_{SUSY})^2}}~,}
 \end{equation}
For completeness, we note that $v^{\DR}(M_Z)$, which is used in the
threshold corrections to the gauge and Yukawa couplings, is
calculated separately via the relation
\begin{equation}
v^{\DR}(M_Z)= \sqrt{{M_Z^{\DR}(M_Z)^2}~\frac{(1- \sin^2\Theta^{\DR}_W)\sin^2\Theta^{\DR}_W}{\pi \alpha^{\DR}} }~.
\end{equation}
In this way, $v^{\DR}(M_Z)$ and $v^{\DR}(M_{SUSY})$
are not strictly correlated via an RGE evolution, but differences of a
higher order are present.

\item \NMSSMTools calculates $v^{\DR}(M_{SUSY})$ via
\begin{equation}
\label{eq:vNMSSMTOOLS}
v^{\DR}(M_{SUSY})^{-2}= \sqrt{2}\,G_\mu\,\left(1
  -\frac{\Pi^T_{WW}(0)}{M_W^2} - \delta_{\rm VB}\right) \;, \\ 
\end{equation}
where $\delta_{\rm VB}$ denotes the one-loop correction to the muon
decay amplitude. In the latter, \NMSSMTools includes only the
  SM contribution.

\item \NMSSMCalc uses also for $v$ an OS condition in terms of the three physical inputs:
 \begin{equation}
 v~=~\frac{2\,M_W}{e}\,\sqrt{1-M_W^2/M_Z^2} ~.
 \end{equation}
\end{enumerate}

\subsubsection{Scales}
\label{sec:scales}

We have so far discussed differences in the
threshold calculations, and  how the codes perform those calculations at different scales. There
are two other important scales involved in the calculation
of the Higgs-boson masses: the scale at which the input
parameters are taken by the codes, and the scale at which the pole
masses are calculated. Thus, for a detailed comparison it is necessary
that these scales be exactly the same in all
codes.  However, as mentioned in
sec.~\ref{sec:OriginalMasses}, it is not possible to unify these two
scales among the codes without modifications in
\NMSSMTools. In the following, we describe how
these two scales are treated in general by the different codes,
and how -- and to what extent -- they can be
fixed to particular numerical values using the provided options in the
input.
\begin{enumerate}
\item \SPheno: If no scale is defined in the input, the input
  parameters are taken at $Q=\sqrt{m_{\tilde t_1} m_{\tilde t_2}}$ and
  the masses are calculated at the same scale. If a scale is set
   via the entries {\tt EXTPAR[0]} or {\tt MODSEL[12]},
  the input parameters are defined at this scale and the masses are
  calculated there as well.

\item \NMSSMCalc uses a similar approach to \SPheno and always
  calculates the masses at the same scale where the input parameters
  are taken. However, if no explicit scale is defined, the default
  choice is $Q=\sqrt{m_{\tilde{t}_L} m_{\tilde t_R}}$.

\item \SoftSUSY by default takes the input scale to be the
  gauge-coupling unification scale.  However, one may instead choose
  the input scale by either explicitly specifying it in {\tt
    EXTPAR[0]} or by setting {\tt EXTPAR[0]} to {\tt -1} in order to
  set the input scale to $Q=\sqrt{m_{\tilde t_1} m_{\tilde t_2}}$. In
  contrast to \SPheno, changing this entry does not affect the scale at
  which the pole masses are calculated.  \SoftSUSY calculates the pole
  masses at $Q=\sqrt{m_{\tilde t_1} m_{\tilde t_2}}$ by default, like
  \SPheno.  To change this one may set an alternative
  scale\footnote{For this scale to be used directly it must be larger
    than $M_Z$.  See documentation for full details.} in the special
  {\tt SOFTSUSY} block, using entry {\tt 4}.

\item \FlexibleSUSY: Both the parameter input scale and the scale at
  which the pole masses are calculated may be chosen by the user in
  the model file used to generate the code.
  In the NMSSM model file used in this section, both scales are set via the entry
  {\tt EXTPAR[0]}.  In the model file used in sec.\ \ref{sec:GUT} to
  study GUT scenarios the parameter input scale is set to be the
  gauge-coupling unification scale, and the scale where the pole
  masses are calculated is set to a generalization of the SUSY scale
  which reduces to $Q=\sqrt{m_{\tilde t_1} m_{\tilde t_2}}$ in the
  absence of family mixing.
  
\item \NMSSMTools takes by default $Q=\sqrt{(2 m^2_{\tilde{u}_L}+
    m^2_{\tilde u_R} + m^2_{\tilde d_R})}/2$ for the input parameters and
  $Q'=\sqrt{m_{\tilde{t}_L} m_{\tilde t_R}}$ to calculate the
  Higgs, stop and sbottom
  pole masses. It is possible to change $Q$ to a fixed value
  via {\tt EXTPAR[0]} or {\tt MINPAR[0]}, but fixing $Q'$ is only possible by modifying
  the code (which is done for the present study).  The running between the two scales is performed
  with two-loop RGEs for the NMSSM. 
\end{enumerate}

\subsubsection{Included two-loop corrections}
While all codes use complete calculations of
the one-loop corrections to the Higgs
masses~\cite{Degrassi:2009yq,Staub:2010ty,Ender:2011qh,Graf:2012hh},
the partial two-loop corrections implemented in the codes
cover different contributions. All the mentioned two-loop
corrections are calculated for zero external momentum.
\begin{enumerate}
\item \NMSSMCalc includes only the two-loop corrections of ${\cal
    O}(\alpha_s\alpha_t)$~\cite{Muhlleitner:2014vsa}, but
  accounts also for the two-loop corrections to the
  electroweak VEV, which are omitted in
    the other codes.
\item \SoftSUSY, \FlexibleSUSY and \NMSSMTools use the 
  two-loop corrections of ${\cal O}(\alpha_s (\alpha_t +
  \alpha_b))$ which were calculated in the context of the NMSSM
  in Ref.~\cite{Degrassi:2009yq}. In addition, these codes use
  the MSSM results for the two-loop corrections of ${\cal
    O}((\alpha_t+\alpha_b)^2)$ ~\cite{Dedes:2003km}. Moreover,
  \SoftSUSY and \FlexibleSUSY include also routines to calculate the
  MSSM results for the tiny two-loop corrections involving
  $\alpha_\tau$~\cite{Brignole:2001jy,Dedes:2002dy,Dedes:2003km,Allanach:2004rh}.
\item \SPheno performs a  two-loop calculation in the gaugeless
  limit. Namely, any two-loop
  correction in the neutral Higgs sector that is
  independent of $g_1$ and $g_2$ is included.  However, \SPheno includes also
  the available routines for the NMSSM corrections of ${\cal
    O}(\alpha_s (\alpha_t + \alpha_b))$, as well as MSSM corrections
  involving only Yukawa couplings similar to \SoftSUSY and
  \FlexibleSUSY. If wished by the user, these routines can be used
  instead of the full calculation.
\end{enumerate}

\begin{table}
\centering
\begin{tabular}{|l|cc|>{\centering\arraybackslash}p{1.5cm} >{\centering\arraybackslash}p{1.5cm}  >{\centering\arraybackslash}p{1.8cm} >{\centering\arraybackslash}p{1.8cm} >{\centering\arraybackslash}p{1.5cm}  |}
\hline
 & \multicolumn{2}{|c}{original} & \multicolumn{5}{|c|}{modified}  \\ 
\hline
 & one-loop & two-loop & one-loop & $\alpha_s \alpha_t$ & $\alpha_s (\alpha_t+\alpha_b)$ & $\alpha_s (\alpha_t+\alpha_b)$ + MSSM & `full'   \\
\hline
\multicolumn{8}{|c|}{TP1} \\
\hline 
{\tt FlexibleSUSY } &{119.48} & {123.55} &{119.73}& {124.82} & {124.82}&{123.81}& --- \\ 
{\tt NMSSMCALC }  &{115.49}& {120.34}&{119.75} &{124.85}& --- & --- & --- \\ 
{\tt NMSSMTOOLS }   & --- & {123.52} &{119.75} & {124.84} & {124.84} &{123.84}& ---  \\ 
{\tt SOFTSUSY }  &{119.75} & {123.84} &{119.75} &{124.84}  & {124.84} &{123.84}& ---  \\ 
{\tt SPHENO }   &{120.69} & {124.84} & {119.75}&{124.84}&{124.84}&{123.84}& {123.84}  \\ 
\hline
\multicolumn{8}{|c|}{TP2} \\
\hline 
{\tt FlexibleSUSY } &{116.28} & {122.84} &{116.46}& {123.79} & {123.79}&{123.06}& --- \\ 
{\tt NMSSMCALC }  &{111.80}& {118.57}&{116.49} &{123.82}& --- & --- & --- \\ 
{\tt NMSSMTOOLS }  & --- & {121.83} &{116.49} & {123.82} & {123.82} &{123.08}& ---  \\ 
{\tt SOFTSUSY }  &{116.49} & {123.08} &{116.49} &{123.82}  & {123.82} &{123.08}& ---  \\ 
{\tt SPHENO }   &{118.00} & {124.74} & {116.49}&{123.81}&{123.81}&{123.08}& {123.05}  \\ 
\hline
\multicolumn{8}{|c|}{TP3} \\
\hline 
{\tt FlexibleSUSY } &{124.16} & {126.58} &{124.15}& {127.55} & {127.55}&{126.59}& --- \\ 
{\tt NMSSMCALC }  &{121.76}& {124.86}&{124.15} &{127.56}& --- & --- & --- \\ 
{\tt NMSSMTOOLS }  & --- & {127.28} &{124.15} & {127.56} & {127.56} &{126.60}& --- \\  
{\tt SOFTSUSY }  &{124.15} & {126.59} &{124.15} &{127.56}  & {127.56} &{126.59}& ---  \\ 
{\tt SPHENO }   &{124.79} & {126.77} & {124.15}&{127.55}&{127.55}&{126.59}& {126.10}  \\ 
\hline 
\multicolumn{8}{|c|}{TP4} \\
\hline 
{\tt FlexibleSUSY } &{124.72} & {127.62} &{124.59}& {128.23} & {128.23}&{127.51}& --- \\ 
{\tt NMSSMCALC }  &{122.85}& {126.37}&{124.59} &{128.23}& --- & --- & --- \\ 
{\tt NMSSMTOOLS }  & --- & {127.30} &{124.59} & {128.24} & {128.24} &{127.52}& ---   \\ 
{\tt SOFTSUSY }  &{124.59} & {127.52} &{124.59} &{128.24}  & {128.24} &{127.52}& ---  \\ 
{\tt SPHENO }   &{124.89} & {126.62} & {124.59}&{128.23}&{128.23}&{127.52}& {126.33}  \\ 
\hline
\multicolumn{8}{|c|}{TP5} \\
\hline 
{\tt FlexibleSUSY } &{122.53} & {125.08} &{122.54}& {126.11} & {126.11}&{125.11}& --- \\ 
{\tt NMSSMCALC }  &{121.95}& {123.14}&{122.54} &{126.12}& --- & --- & --- \\ 
{\tt NMSSMTOOLS }  & --- & {126.95} &{122.54} & {126.12} & {126.12} &{125.12}& --- \\ 
{\tt SOFTSUSY }  &{122.54} & {125.12} &{122.54} &{126.12}  & {126.12} &{125.12}& ---  \\ 
{\tt SPHENO }   &{122.90} & {125.61} & {122.53}&{126.12}&{126.12}&{125.12}& {124.85}  \\ 
\hline 
\multicolumn{8}{|c|}{TP6} \\
\hline 
{\tt FlexibleSUSY } &{121.92} & {126.46} &{122.09}& {127.7} & {127.7}&{126.64}& --- \\ 
{\tt NMSSMCALC }  &{118.55}& {123.45}&{122.12} &{127.73}& --- & --- & --- \\ 
{\tt NMSSMTOOLS }  & --- & {126.63} &{122.12} & {127.73} & {127.73} &{126.67}& ---  \\ 
{\tt SOFTSUSY }  &{122.12} & {126.67} &{122.12} &{127.73}  & {127.73} &{126.67}& ---  \\ 
{\tt SPHENO }   &{123.37} & {131.29} & {122.09}&{127.71}&{127.71}&{126.65}& {129.91}  \\ 
\hline
\end{tabular} 
\caption{The mass in \gev\ of the SM-like scalar after applying the modifications listed in sec.~\ref{sec:modifications} for TP1--TP6 at the different loop levels and including different two-loop corrections. All two-loop corrections are for zero external momentum, while at one-loop the entire $p^2$ dependence is taken into account. ``full'' refers to a calculation which includes all corrections in the gaugeless limit where electroweak gauge couplings are neglected. Since the original version of \NMSSMTools does not provide a flag to turn off the two-loop corrections,  we don't give any one-loop value here. }
\label{tab:adjustedMasses}
\end{table}

\subsection{Adjusting the codes}
\label{sec:modifications}
To check if the differences between the
  codes listed in the previous section can really explain the
sizable discrepancies in the Higgs-mass
predictions,  we have modified the codes,
choosing \SoftSUSY as
reference. This choice resulted from
  mere convenience: \FlexibleSUSY is very close to \SoftSUSY, and the
differences between the two codes were already
fully understood. This made it easy to adapt \FlexibleSUSY and left
just three codes which required more
modifications. In particular, we have changed the following:
\begin{itemize}
\item {\bf Calculation of the \DR parameters}: Changing the codes to
  mimic the calculation of the running couplings of \SoftSUSY would
  be, of course, a big task beyond the scope of this
  comparison\footnote{We stress that, in
        the context of this comparison, we double-checked
       the corresponding
      calculations performed by the codes and did not find any bugs.}
  Therefore, the chosen approach was to read in directly the values of
  $g_i$, $Y_i$, $v$ and $\tan\beta$ at the scale $Q$ as
  calculated by \SoftSUSY and use them in the calculation of the Higgs
  masses. This was done for \NMSSMTools, \NMSSMCalc and \SPheno.  In
  \FlexibleSUSY no actual modifications to the code are made, but we
  will show results for both options for calculating the Weinberg
  angle.  The option which matches the approach in \SoftSUSY will be
  treated as the modified version, while the approach implemented in
  \FlexibleSUSY 1.0.0 will be treated as the original version.
\item {\bf Renormalization scale}: \NMSSMTools was forced to calculate
  the Higgs mass at the same scale as all other codes.
\item {\bf Renormalization scheme}: The finite parts of the
  counterterms of the $W$ boson mass $\delta M_W$, the $Z$ boson mass
  $\delta M_Z$, and the electric charge $\delta e$ were put to zero in
  \NMSSMCalc. Note that at two-loop level it is more convenient to
  introduce a counterterm for $v$, since in the gaugeless limit  $\delta M_Z^2$ and $\delta M_W^2$ are zero, 
  however, $\delta M_Z^2/M_Z^2$ and $\delta M_W^2/M_W^2$ not. 
  The remaining combinations can be expressed by $\delta v$.
   This counterterm
  also needs to be set to zero to adjust to the other codes.
\item {\bf Loop corrections}: Of course no additional loop corrections
  have been implemented in any of the codes in the context of this
  analysis. However, we added options to the different codes to turn
  off specific loop corrections, to get a clear picture of the masses
  at one-loop and at two-loop including (i) ${\cal O}(\alpha_s
  \alpha_t)$, (ii) ${\cal O}(\alpha_s
  (\alpha_t+\alpha_b))$ (iii) ${\cal O}(\alpha_s
  (\alpha_t+\alpha_b))$ plus MSSM
  approximations for the corrections involving only Yukawa
    couplings, (iv) the full NMSSM calculation in the gaugeless limit.
\end{itemize}

\subsection{Masses after the adjustments}
The results for the SM-like Higgs mass after applying all changes to
the codes are listed in Tab.~\ref{tab:adjustedMasses}. Here, we
disentangled also the different loop corrections at the one- and
two-loop level. The entries $|Z_{i1}^H|$ and $|Z_{i3}^H|$ of the
corresponding rotation matrix are listed in
Tab.~\ref{tab:adjustedZH}.

\begin{table}
\centering
\begin{tabular}{|l|cc|>{\centering\arraybackslash}p{1.75cm} >{\centering\arraybackslash}p{1.75cm} >{\centering\arraybackslash}p{1.75cm} >{\centering\arraybackslash}p{1.8cm}  |}
\hline
 & \multicolumn{2}{|c}{original} & \multicolumn{4}{|c|}{modified}  \\ 
\hline
 & one-loop & two-loop & one-loop & $\alpha_s \alpha_t$ & $\alpha_s (\alpha_t+\alpha_b)$ + MSSM & `full'   \\
\hline
\multicolumn{7}{|c|}{TP1} \\
\hline 
{\tt FlexibleSUSY } &{0.1039,0.0076} & {0.1039,0.0076} &{0.1039,0.0076}& {0.1039,0.0076} & {0.1039,0.0076}& --- \\ 
{\tt NMSSMCALC }  &{0.1039,0.0076}& {0.1039,0.0076}&{0.1039,0.0076} &{0.1039,0.0076}& ---  & --- \\ 
{\tt NMSSMTOOLS }  & --- & {0.1039,0.0076} &{0.1039,0.0076} & {0.1039,0.0076} & {0.1039,0.0076}& ---  \\ 
{\tt SOFTSUSY }  &{0.1039,0.0076} & {0.1039,0.0076} &{0.1039,0.0076} &{0.1039,0.0076}   &{0.1039,0.0076}& ---  \\ 
{\tt SPHENO }   &{0.1040,0.0076} & {0.1040,0.0074} & {0.1039,0.0076}&{0.1039,0.0076}&{0.1039,0.0076}& {0.1039,0.0075}  \\ 
\hline
\multicolumn{7}{|c|}{TP2} \\
\hline 
{\tt FlexibleSUSY } &{0.1034,0.0004} & {0.1034,0.0004} &{0.1034,0.0004}& {0.1034,0.0004} &{0.1034,0.0004}& --- \\ 
{\tt NMSSMCALC }  &{0.1034,0.0004}& {0.1034,0.0004}&{0.1034,0.0004} &{0.1034,0.0004}&  --- & --- \\ 
{\tt NMSSMTOOLS }  & --- & {0.1038,0.0004} &{0.1034,0.0004} & {0.1034,0.0004} & {0.1034,0.0004}& ---  \\ 
{\tt SOFTSUSY }  &{0.1034,0.0004} & {0.1034,0.0004} &{0.1034,0.0004} &{0.1034,0.0004}   &{0.1034,0.0004}& ---  \\ 
{\tt SPHENO }   &{0.1035,0.0004} & {0.1035,0.0004} & {0.1034,0.0004}&{0.1034,0.0004}&{0.1034,0.0004}& {0.1034,0.0004}  \\ 
\hline
\multicolumn{7}{|c|}{TP3} \\
\hline 
{\tt FlexibleSUSY } &{0.2687,0.2975} & {0.2813,0.2470} &{0.2693,0.2970}& {0.2835,0.2400} &{0.2819,0.2465}& --- \\ 
{\tt NMSSMCALC }  &{0.2664,0.3081}& {0.2811,0.2502}&{0.2705,0.2935} &{0.2842,0.2385} & --- & --- \\ 
{\tt NMSSMTOOLS }  & --- & {0.2782,0.2708} &{0.2705,0.2935} & {0.2842,0.2384} & {0.2828,0.2442}& ---  \\ 
{\tt SOFTSUSY }  &{0.2693,0.2970} & {0.2820,0.2460} &{0.2693,0.2970} &{0.2836,0.2399}  &{0.2820,0.2460}& ---  \\ 
{\tt SPHENO }   &{0.2701,0.2951} & {0.2725,0.2851} & {0.2693,0.2969}&{0.2836,0.2398}&{0.2820,0.2459}& {0.2719,0.2863}  \\ 
\hline
\multicolumn{7}{|c|}{TP4} \\
\hline 
{\tt FlexibleSUSY } &{0.4852,0.3689} & {0.4866,0.3712} &{0.4853,0.3698}& {0.4862,0.3938}&{0.4868,0.3722}& --- \\ 
{\tt NMSSMCALC }  &{0.4858,0.3625}& {0.4869,0.3832}&{0.4845,0.4098} &{0.4836,0.4476}&  --- & --- \\ 
{\tt NMSSMTOOLS }  & --- & {0.4855,0.4501} &{0.4845,0.4098} & {0.4837,0.4472} & {0.4851,0.4235}& ---  \\ 
{\tt SOFTSUSY }  &{0.4853,0.3700} & {0.4867,0.3725} &{0.4853,0.3700} &{0.4862,0.3941}   &{0.4867,0.3725}& ---  \\ 
{\tt SPHENO }   &{0.4852,0.3810} & {0.4835,0.4262} & {0.4853,0.3698}&{0.4862,0.3939}&{0.4867,0.3723}& {0.4840,0.4133}  \\ 
\hline
\multicolumn{7}{|c|}{TP5} \\
\hline 
{\tt FlexibleSUSY } &{0.0171,0.8787} & {0.2189,0.4681} &{0.0205,0.8754}& {0.2445,0.3881} &{0.2191,0.4685}& --- \\ 
{\tt NMSSMCALC }  &{0.0154,0.9162}& {0.2055,0.5099}&{0.1259,0.7020} &{0.2621,0.3276}&  --- & --- \\ 
{\tt NMSSMTOOLS }  & --- & {0.1974,0.5396} &{0.1259,0.7020} & {0.2621,0.3276} & {0.2485,0.3758}& ---  \\ 
{\tt SOFTSUSY }  &{0.0208,0.8749} & {0.2195,0.4679} &{0.0208,0.8749} &{0.2447,0.3876}  &{0.2195,0.4679}& ---  \\ 
{\tt SPHENO }   &{0.0605,0.8192} & {0.1958,0.5372} & {0.0206,0.8752}&{0.2449,0.387}&{0.2197,0.4672}& {0.1806,0.5766}  \\ 
\hline
\multicolumn{7}{|c|}{TP6} \\
\hline 
{\tt FlexibleSUSY } &{0.2741,0.1946} & {0.2741,0.1950} &{0.2746,0.1937}& {0.2747,0.1942} &{0.2747,0.1941}& --- \\ 
{\tt NMSSMCALC }  &{0.2742,0.1947}& {0.2746,0.1947}&{0.2748,0.1935} &{0.2749,0.1939}&  --- & --- \\ 
{\tt NMSSMTOOLS }  & --- & {0.2743,0.1953} &{0.2748,0.1935} & {0.2749,0.1939} & {0.2749,0.1939}& ---  \\ 
{\tt SOFTSUSY }  &{0.2746,0.1937} & {0.2747,0.1941} &{0.2746,0.1937} &{0.2747,0.1941}   &{0.2747,0.1941}& ---  \\ 
{\tt SPHENO }   &{0.2748,0.1938} & {0.2712,0.2039} & {0.2746,0.1937}&{0.2747,0.1942}&{0.2747,0.1941}& {0.2711,0.2037}  \\ 

\hline
\end{tabular} 
\caption{Absolute value of the $H_d$ and singlet components
  ($|Z^H_{i1}|$, $|Z^H_{i3}|$) of the SM-like Higgs with index $i$
  after the modifications listed in sec.~\ref{sec:modifications} for
  TP1--TP6. Because of the size of the table, we do not show explicitly the 
  largest entry $|Z^H_{i2}|$, but it can easily be calculated as 
  $|Z^{H}_{i2}|^2=1-|Z^{H}_{i1}|^2-|Z^{H}_{i3}|^2$.
  The remaining differences are caused by the external momentum $p^2$
   used to calculate $Z^H$: \NMSSMCalc\ and \NMSSMTools\ 
  set
  $p^2=0$, while the other codes set $p^2=m_{h_1}^2$. The conventions 
  for the given corrections are the same as in
  Tab.~\ref{tab:adjustedMasses}, but we do not show the results for 
  $\alpha_s(\alpha_b+\alpha_t)$ explicitly because they fully agree with the ones for $\alpha_s \alpha_t$ for the number of 
  digits used here.}
  \label{tab:adjustedZH}
\end{table}

The big differences that were observed
before are totally gone when comparing equivalent calculations. Thus,
for TP1--TP5 all codes show an impressive
agreement at ${\cal O}(\alpha_s \alpha_t)$. The corrections of
${\cal O}(\alpha_s \alpha_b)$ are completely negligible for all
considered points (they could, however, give
a sizable effect for very large $\tan\beta$). When including
the purely Yukawa corrections at the two-loop level in the MSSM
approximation, which is possible in all codes but \NMSSMCalc, very
good agreement is also found among the different tools. However, this
approximation might not always be very good, and fails in particular
for large   $\lambda$. This is
  manifest in the comparison with the full NMSSM result, which is
  available only for \SPheno.

\subsection{Size of the different effects}
\label{sec:size}
We discuss now in turn the impact of the different
changes we applied.

\begin{enumerate}
\item {\FlexibleSUSY}: The entire difference between \FlexibleSUSY and
  \SoftSUSY comes from the calculation of the Weinberg angle
  $\sin^{\DR} \Theta_W$ at the scale $M_Z$, as described in
  Section \ref{subsec:calculation-of-drbar-parameters}.

\item {\SPheno}: In Tab.~\ref{tab:spheno} the effects of
adjusting successively the Yukawa
  couplings, the gauge couplings, and the electroweak VEV
  are shown. One can see that the biggest change
  comes from the Yukawa couplings. Of course, the top Yukawa
  coupling plays the main role.
  
\begin{table}[hbt]
\begin{center}
\begin{tabular}{|c|c|ccc|c|}
\hline
Point & original & $Y$  & $g$ & $v$ & modified \\ 
\hline
TP1 & {124.84} & {123.65} & {123.61} & {123.84} & {123.84}  \\ 
TP2 & {124.74} & {123.18} & {123.13} & {123.05} & {123.05}  \\  
TP3 & {126.77} & {126.06} & {126.00} & {126.10} & {126.10}  \\ 
TP4 & {126.62} & {126.21} & {126.16} & {126.33} & {126.33}  \\ 
TP5 & {125.61} & {124.89} & {124.84} & {124.85} & {124.85}  \\ 
TP6 & {131.29} & {130.06} & {130.01} & {129.91} & {129.91}  \\ 
\hline
\end{tabular} 
\end{center}
\caption{The Higgs prediction for the SM-like Higgs mass by 
  \SPheno after applying successively the different adjustments for the 
  Yukawas ($Y$), gauge couplings ($g$), and the electroweak VEV ($v$). 
  Here, ``original'' refers to the results when using the code without any
  modification, while for ``modified'' all adjustments are turned on.}
\label{tab:spheno}
\end{table}

\item {\NMSSMCalc}: Tab.~\ref{tab:nmssmcalc} shows the effects
  of applying successively the different
  adjustments. Here we show the effect of changing
  the Yukawa couplings, changing the renormalization of the EW sector,
  and neglecting ${\cal
      O}(\alpha_s\alpha_t)$ contributions due to the 
  conversion of the electroweak VEV to the \DR parameter, which
    were omitted in the other spectrum calculators. The effect
  of the latter on the Higgs masses is proportional to
  $\lambda$, and hence more visible for
  large $\lambda$.

\begin{table}[hbt]
\begin{center}
\begin{tabular}{|c|c|ccc|c|}
\hline
Point & original & $Y$  & $\delta_1$ &$\delta_2$ &modified \\ 
\hline
TP1 & {120.34} & {124.41}  & {124.85} & {124.85} & {124.85} \\ 
TP2 & {118.57} & {123.31}  & {123.82} & {123.82} & {123.82} \\ 
TP3 & {124.86} & {127.55}  & {127.50} & {127.56} & {127.56}  \\ 
TP4 & {126.37} & {128.32}  & {128.18} & {128.23} & {128.23}  \\ 
TP5 & {123.14} & {126.21}  & {126.03} & {126.12} & {126.12}  \\ 
TP6 & {123.45} & {127.26}  &{127.55} &{127.73} & {127.73}  \\ 
\hline
\end{tabular}
\end{center}
\caption{The prediction for the SM-like Higgs mass by
  \NMSSMCalc\ after adjusting 
  the \DR Yukawa
  couplings ($Y$); performing a \DR\ renormalization of the 
  EW sector by changing the values of  $g_1$, $g_2$ and $v$ and removing
  the finite parts of the one-loop counter-terms $\delta M_W$, $\delta M_Z$ and
  $\delta e$ ($\delta_1$); removing also
  the finite counterterm for $v$ at the two-loop level ($\delta_2$). The same conventions as 
  for Tab.~\ref{tab:spheno} are used. }
\label{tab:nmssmcalc}
\end{table}

\item {\NMSSMTools}:   Tab.~\ref{tab:nmssmtools} shows the effects
  of applying successively the different
  adjustments: the effects of adjusting
  successively the scale $Q'$ for the Higgs-mass calculation,
  the running Yukawa and gauge couplings and the running
  electroweak VEV. The changes in the scale have
    only an effect for TP2, because for the other points $Q$ was
    chosen as $\sqrt{m_{\tilde t_L} m_{\tilde t_R}}$, which coincides with
    the scale $Q'$ used by
    \NMSSMTools.  The only sizable effects are due
  to the top Yukawa coupling and the changes in the VEV.

\begin{table}[hbt]
\begin{center}
\begin{tabular}{|c|c|ccccc|c|}
\hline
Point & original  & $Q'=Q$&   $Y$ &  $g$ & $v$  & modified \\ 
\hline
TP1 & {123.52} & {123.52} & {123.96} & {123.99} & {123.84} & {123.84}  \\ 
TP2 & {121.83} & {121.44} & {123.46} & {123.52} & {123.08} & {123.08}  \\ 
TP3 & {127.28} & {127.28} & {127.43} & {127.43} & {126.60} & {126.60}  \\ 
TP4 & {127.30} & {127.30} & {127.13} & {127.07} & {127.52} & {127.52}  \\ 
TP5 & {126.95} & {126.95} & {127.34} & {127.45} & {125.12} & {125.12}  \\ 
TP6 & {126.63} & {126.63} & {127.56} & {127.66} & {126.67} & {126.67}  \\ 
\hline
\end{tabular} 
\end{center}
\caption{Changes in 
  the prediction by \NMSSMTools for the SM-like Higgs mass 
  after forcing the SUSY scale and the scale for the mass calculation to be
  identical ($Q'=Q$), changing the Yukawa couplings ($Y$), the gauge
  couplings ($g$), and the electroweak VEV ($v$). The same conventions
  as for Tab.~\ref{tab:spheno} are used. 
}
\label{tab:nmssmtools}
\end{table}
\end{enumerate}

One can see from this comparison that the main source of discrepancies between the predictions of the five codes is the \DR{-renormalized} top Yukawa coupling used in the calculation of the Higgs masses: differences in the determination of $Y^{\DR}_t(Q)$ have a larger impact than additional two-loop corrections to the Higgs mass matrix. In particular, the fact that \NMSSMCalc includes only (S)QCD corrections in the calculation of $Y^{\DR}_t(Q)$ results in a prediction for the SM-like Higgs mass a few GeV below the ones of the other codes. 
The results in tables \ref{tab:adjustedMasses} and \ref{tab:nmssmcalc} suggest that the inclusion of additional one-loop corrections in the calculation of that coupling by \NMSSMCalc would account for the bulk of the corresponding two-loop corrections to the Higgs masses, thus reducing the discrepancies with the other codes. However, the precise calculation of the Higgs decays in \NMSSMCalc includes only (S)QCD corrections. Therefore, with the aim of preserving the overall consistency of the calculation, the authors of \NMSSMCalc include only (S)QCD corrections in $Y^{\DR}_t(Q)$ as well.

\section{Expected effects for GUT scenarios}
\label{sec:GUT}
We have focused so far on the Higgs-mass
calculation in which all SUSY parameters
are fixed by the input at the SUSY scale. However, most 
codes offer also the possibility to study a GUT scenario where
the soft SUSY-breaking parameters are fixed at the GUT scale. The
 differences among the codes discussed in the
  previous sections are going to play an important role
here as well.  Indeed, because of the additional RGE running, the
discrepancies between the results of the
  different codes are usually of the same size as, and often
even bigger than those found in the case of SUSY-scale
  inputs with comparable parameter choices.  
  In particular, the differences in the determination of the \DR couplings can play
a crucial role, because the different values for the 
couplings at the SUSY scale affect the entire running
between $M_{SUSY}$ and $M_{GUT}$. This is especially important for
points in which the input value of $\lambda$
at $M_{SUSY}$ is close to the perturbativity limit.  Small
changes in the threshold corrections to the weak-scale
  couplings can change the value of $\lambda$ at the GUT scale
considerably.  This will then change the running of all soft terms
from the GUT scale to the SUSY scale. While the modifications
in the sfermion parameters are not necessarily large, $A_\lambda$ and
$A_\kappa$ can be strongly affected. This will then have an impact on
the singlet-like scalar.

To demonstrate this effect, we consider the two points given in
Tab.~\ref{tab:inputGTP}. Both points provide an SM-like Higgs mass in
the preferred range together with a light singlet,
but the first (GTP1) is characterized by small values of
  $\lambda$ and $\kappa$, while the second (GTP2) is characterized by
  larger values of those couplings, not far below the perturbativity
  limit. To show the large sensitivity of GTP2 compared to GTP1 to
the parameters at the SUSY scale, we vary the top mass within the
uncertainty: $m_t = 173.2 \pm 1.5~\gev$. The masses of the three
CP-even Higgs bosons computed by the four codes that allow
  for GUT-scale boundary conditions, i.e.~\FlexibleSUSY, \NMSSMTools,
  \SoftSUSY and \SPheno, are shown in Tab.~\ref{tab:resGTP}. We 
  have used for this comparison the un-tuned versions of all codes.

\begin{table}[hbt]
\centering 
\begin{tabular}{|c|ccc|cccccc|}
\hline 
 & $m_0$ & $M_{1/2}$ &  $A_0$ & $\tan\beta$ & $\lambda$ & $\kappa$ & $A_\lambda$ & $A_\kappa$ & $\mu_{\rm eff} $ \\
\hline 
GTP1 &  1000 & 1000 & -3000 & 10 & 0.05 & 0.05 & -150 & -300 & 100 \\[1mm]
GTP2 &  1000 & 1000 & -3000 & 10 & 0.48 & 0.46 & -150 & -300 & 100 \\
\hline  
\end{tabular}
\caption{Input values for GTP1 and GTP2. All dimensionful parameters in units of \gev. The parameters in the left section of the table are defined at the GUT scale, whereas those in the right section are defined at the scale $M_{SUSY}$, with the exception of $\tan\beta$ which is defined at $M_Z$.}
\label{tab:inputGTP}
\end{table}

\begin{table}[hbt]
\centering
\begin{tabular}{|c|ccc|}
\hline 
              &  $ m_t = 171.7~\gev$ &  $ m_t = 173.2~\gev$ &  $ m_t = 174.7~\gev$ \\
\hline 
\FlexibleSUSY &  100.1,\ 122.2, \ 951.9 & 100.0,\ 123.3,\ 967.0 & 100.0,\ 124.5,\ 982.0 \\
\NMSSMTools   &  100.0,\ 119.3, \ 982.1 & 100.0,\ 120.1,\ 974.4 &  99.9,\ 120.8,\ 1014.0 \\
\SoftSUSY     &  100.1,\ 122.5, \ 958.6 & 100.0,\ 123.6,\ 974.4 & 100.0,\ 124.8,\ 990.2 \\
\SPheno       &  100.0,\ 123.7, \ 976.6 & 99.9, \ 124.9,\ 992.3 & 99.9,\ 126.0,\ 1008.0 \\
\hline 
\end{tabular}

\begin{tabular}{|c|ccc|}
\hline 
              &     $ m_t = 171.7~\gev$ &  $ m_t = 173.2~\gev$ &  $ m_t = 174.7~\gev$ \\
\hline 
\FlexibleSUSY &  102.8,\ 123.3, \ 961.9 & 97.9,\ 124.6,\ 974.1   & 92.5,\ 126.1,\ 986.1 \\
\NMSSMTools   &  96.6,\ 118.8, \ 987.7  & 91.4,\ 119.1,\ 1000.8  &  no output \\
\SoftSUSY     &  102.4,\ 123.7, \ 968.2 & 97.3,\ 125.2,\ 981.1   & 91.4,\ 126.7,\ 994.0 \\
\SPheno       &  95.1,\ 125.3, \ 981.9  & 89.3, \ 126.8,\ 994.6  & 82.5,\ 128.3,\ 1007.5 \\
\hline 
\end{tabular}
\caption{The masses of the three CP-even Higgs bosons, $m_{h_{1,2,3}}$, for GTP1 (top) and GTP2 (bottom) for three different values of the top quark mass, $m_t$.}
\label{tab:resGTP}
\end{table}

We see that, for GTP1, the SM-like Higgs mass varies by about 2~GeV
for all codes when $m_t$ is varied within one standard
  deviation, while the light-singlet mass is 
fairly independent of the top mass,  and
the predictions of all codes for the light-singlet
  mass agree within 0.1~\gev. For GTP2, the variation with
  $m_t$ in the SM-like Higgs mass is about 3~GeV, but the
singlet is very sensitive to the value
of the top mass: its mass can change by about 10 GeV, and  the different
codes predict masses that
differ from each other by several \gev. In addition,
for $ m_t = 174.7$~GeV \NMSSMTools runs into
the Landau pole and is not 
able to calculate the spectrum. The situation can be even more
dramatic when taking for instance the benchmark points proposed in
Ref.~\cite{Beskidt:2013gia}: for each of the three points BP1--BP3, at
least two codes do not produce an output. When they are forced to
produce an output (for instance by ignoring tachyonic states which
appear at tree-level), the differences between the codes can reach
20--30~GeV for the light singlet.

Points with very large $\lambda$ at the SUSY scale are
attractive since they seem to soften the SUSY fine-tuning
problem \cite{BasteroGil:2000bw,Dermisek:2005gg,Ellwanger:2011mu,Ross:2011xv,Ross:2012nr,Gherghetta:2012gb,Perelstein:2012qg,Kim:2013uxa,Kaminska:2014wia,Binjonaid:2014oga}. However, if $\lambda$ is too close to the perturbativity
limit, the output produced by the spectrum generators has to be taken
with some care: these points can be very sensitive to small variations
of the input parameters. Changing the SM input parameters within their
respective uncertainties -- or changing the details of the
  determination of the running couplings -- can lead to a completely
different spectrum, or even result in the spectrum generator classifying the parameter point as unphysical.

\section{Conclusions}
\label{sec:conclusion}

We have discussed the differences in the predictions for the
scalar Higgs masses and mixings by public spectrum generators
for the NMSSM. It has been shown that sizeable
discrepancies can show up for various
parameter points, but the origins of those discrepancies
 are fully understood. The main sources of discrepancies are in the
approach used to calculate the running parameters that
enter the Higgs-mass calculation. The most important parameter
here is the top Yukawa coupling. Further
  discrepancies arise from the different accuracy of the two-loop
  corrections included in the codes.

This has important consequences for phenomenological
studies:  For MSSM-like points, \FlexibleSUSY,
\NMSSMTools, \SoftSUSY and \SPheno usually agree quite well
with each other, while \NMSSMCalc returns in general a smaller
value for the SM-like Higgs boson mass. The main
reasons for this discrepancy are the value of $Y^{\DR}_t$ at $M_{SUSY}$, which
is always smaller in \NMSSMCalc, and the inclusion of the MSSM results for the ${\cal
  O}((\alpha_t+\alpha_b)^2)$ corrections in \FlexibleSUSY, \NMSSMTools
and \SoftSUSY but not in \NMSSMCalc. For MSSM-like points, the MSSM results
provide a good
  approximation to the  calculation implemented in \SPheno\
  which includes the NMSSM specific effects.  
  On the other hand, this approximation usually fails for large values of
$\lambda$ and/or if light singlet states are present. In these cases,
the results of \FlexibleSUSY, \NMSSMTools and \SoftSUSY have to be
taken with care because it might be that the MSSM approximation has
even the wrong sign compared to the full NMSSM calculation
\cite{Goodsell:2014pla}. 
For these cases \SPheno, with the additional NMSSM specific contributions, is expected to provide the most reliable prediction, with the caveat that if states in the loop become very light,  the calculation can be plagued by intrinsic problems of the effective-potential calculation \cite{Martin:2013gka,Elias-Miro:2014pca,Martin:2014bca}.
\NMSSMCalc, on the other hand,
  is the only code that includes two-loop
  ${\cal O}(\alpha_t\alpha_s)$ corrections to $v$, which
  become non-negligible for
    large values of $\lambda$.
    
 It has been demonstrated with one example, that the differences can be much more pronounced
for GUT scenarios with values of $\lambda$ close to the perturbativity limit. The differences in the 
threshold calculations among the codes can cause huge discrepancies in particular for light singlets and the predicted 
masses can differ by 10~\gev\ and more.

In general, comparing the
  predictions of different codes may provide a ballpark indication for the
theoretical uncertainty of the Higgs mass
  calculation. However, even if these codes agree very well, one
should  not necessarily assume that the
 uncertainty is small: it is known from the MSSM that spectrum 
 generators performing a \DR\ calculation (\SoftSUSY, \SPheno, {\tt Suspect} \cite{Djouadi:2002ze})
 can agree quite well, while 
 sizeable differences to the OS calculation of {\tt FeynHiggs} exists.
 The differences are assumed to come from the missing 
 electroweak corrections and momentum dependence at two-loop level as well as from the 
 dominant three-loop corrections.  These are
the effects that
  underlie the often-quoted estimate of a 3~GeV uncertainty for
  the SM-like Higgs mass in the
  MSSM~\cite{Allanach:2004rh,Degrassi:2002fi}. To get an estimate of
  the remaining theoretical uncertainty in the NMSSM, a comparison 
between different renormalization schemes will  therefore be
necessary. This is deferred to future work, when
  further publicly available codes will have implemented OS renormalization.
 However, even the validity
  of such an estimate is under discussion already in the MSSM, in particular for a heavier
SUSY scale of 2 TeV and larger. Calculations based on the effective
field theory approach, which become valid in
this mass range, usually predict a SM-like Higgs mass that is
 lower by a few GeV compared to the prediction from dedicated tools for the MSSM \cite{Draper:2013oza,Bagnaschi:2014rsa,Vega:2015fna}. 

With this work an important first step has been taken to understand
the large differences in the Higgs mass computations of the various
public NMSSM codes. For the investigated codes these differences are
now completely understood and it has been shown that discrepancies of more
than 3~GeV are not unusual taking into account the different
corrections implemented. To further pin down the residual theoretical
uncertainty the next natural step will be to 
extend this investigation to the comparison of different
renormalization schemes.

\section*{Acknowledgements}
We thank Peter Drechsel, Mark D. Goodsell, Werner Porod,
Heidi Rzehak,
  Michael Spira, Kathrin Walz
and Georg Weiglein for many useful discussions.

This work was supported in part by the Research Executive Agency (REA)
of the European Commission under the Initial Training Networks
``INVISIBLES'' (PITN-GA-2011-289442) and ``Higgs-Tools''
(PITN-GA-2012-316704), and by the European Research Council (ERC)
under the Advanced Grant ``Higgs@LHC''
(ERC-2012-ADG\_20120216-321133).  The work of P.~S.~at LPTHE is
supported in part by French state funds managed by the ANR
(ANR-11-IDEX-0004-02) in the context of the ILP LABEX
(ANR-10-LABX-63). The work of P.~A.~was supported by the ARC Centre of
Excellence for Particle Physics at the Terascale.

\begin{appendix}
\lstset{
	keywordstyle=\bfseries,
	showstringspaces=false,
        prebreak=\raisebox{0ex}[0ex][0ex] {\ensuremath{\hookrightarrow}},
        postbreak=\raisebox{0ex}[0ex][0ex]{\ensuremath{\space}},
        breaklines=true, 
        breakatwhitespace=true,
        frame=none
}

\lstdefinestyle{file}{
        basicstyle=\ttfamily\mdseries,
	language=bash,
	frame=shadowbox,
        numbers=left,   
        numberstyle=\tiny}

\section{Used options in input files}

\subsection{\FlexibleSUSY}
\begin{lstlisting}[style=file]
Block MODSEL                 # Select model
    6   0                    # flavour violation
   12   SCALE		     #
Block FlexibleSUSY
    0   1.000000000e-04      # precision goal
    1   0                    # max. iterations (0 = automatic)
    2   0                    # algorithm (0 = two_scale, 1 = lattice)
    3   0                    # calculate SM pole masses
    4   2                    # pole mass loop order
    5   2                    # EWSB loop order
    6   2                    # beta-functions loop order
    7   1                    # threshold corrections loop order
    8   1                    # Higgs 2-loop corrections O(alpha_t alpha_s)
    9   1                    # Higgs 2-loop corrections O(alpha_b alpha_s)
   10   1                    # Higgs 2-loop corrections O(alpha_t^2 + alpha_t alpha_b + alpha_b^2)
   11   1                    # Higgs 2-loop corrections O(alpha_tau^2)
\end{lstlisting}

\subsection{\NMSSMCalc}
\begin{lstlisting}[style=file]
Block MODSEL
  3  1   # NMSSM
  5  0   # 0: CP-conserving; 2: general CP-violation
  6  2   # loop level 1: one 2:two
  7  1   # 1: DRbar scheme for top/stop-sector; 2: OS scheme for top/stop-sector
\end{lstlisting}

\subsection{\NMSSMTools}
\begin{lstlisting}[style=file]
BLOCK MODSEL
	3	1		# NMSSM particle content
	1	0		# IMOD (0=general NMSSM, 1=mSUGRA, 2=GMSB)
	10	0		# ISCAN (0=no scan, 1=grid scan, 2=random scan, 3=MCMC)
	9	0		# |OMGFLAG|=0: no (default), =1: relic density only,
#				  =2: dir. det. rate, =3: indir. det. rate, =4: both,
#				  OMGFLAG>0: 0.107<OMG<0.131, <0: OMG<0.131
	8       2               # Precision for Higgs masses (default 0: as before,
#                                 1: full 1 loop + full 2 loop from top/bot Yukawas
#				  2: as 1 + pole masses - 1&2 by courtesy of P. Slavich)
	13      0               # 1: Sparticle decays via NMSDECAY (default 0)
	11      0               # Constraints on (g-2)_muon (1=yes, 0=no, default=1)
	14      0               # 0: H-> VV,VV* (default); 1: H->VV,VV*,V*V*
	15	0		# Precision for micromegas (defalt=0):
#				  +0/1: fast computation on/off
#				  +0/2: Beps=1d-3, 1d-6
#				  +0/4: virtual Ws off/on
#

BLOCK MINPAR
 	0	SCALE   # MSUSY (If =/= SQRT(2*MQ1+MU1+MD1)/2)
#
\end{lstlisting}

\subsection{\SoftSUSY}
\begin{lstlisting}[style=file]
# Example NMSSM input in SLHA2 format
Block MODSEL                 # Select model
    6    0                   # flavour violation
    1    0                   # mSUGRA
    3    1                   # NMSSM
   12  SCALE                 #
Block SOFTSUSY               # SOFTSUSY specific inputs
    1   1.000000000e-04      # tolerance
    2   2.000000000e+00      # up-quark mixing (=1) or down (=2)
    5   1.000000000E+00      # 2-loop running
    3   1.000000000E+00      # printout
        4   SCALE                #
    7   4.0		
   15   1.000000000E+00      # NMSSMTools compatible output (default: 0)
   16   0.000000000E+00      # Call micrOmegas (default: 0 = no,
                             # 1 = relic density only,
                             # 2 = direct detection + relic density,
                             # 3 = indirect detection + relic density,
                             # 4 = all)
   17   1.000000000E+00      # Sparticle decays via NMSDECAY (default: 0)
   18   1.000000000E+00      # use soft Higgs masses as EWSB output
 
\end{lstlisting}

\subsection{\SPheno}
\begin{lstlisting}[style=file]
Block MODSEL      #  
1 0               #  1/0: High/low scale input 
2 1              # Boundary Condition  
6 0               # Generation Mixing 
12  SCALE     #
Block SPhenoInput   # SPheno specific input 
  1 -1              # error level 
  2  0              # SPA conventions 
  7  0              # Skip 2-loop Higgs corrections 
  8  3              # Method used for two-loop calculation 
  9  1              # Gaugeless limit used at two-loop 
 10  0              # safe-mode used at two-loop 
 11 0               # calculate branching ratios 
 13 1               # include 3-Body decays 
 12 1.000E-04       # write only branching ratios larger than this value 
 15 1.000E-30       # write only decay if width larger than this value 
 31 -1              # fixed GUT scale (-1: dynamical GUT scale) 
 32 0               # Strict unification 
 34 1.000E-04       # Precision of mass calculation 
 35 80              # Maximal number of iterations
 36 40              # Minimal number of iterations before discarding point
 37 1               # Set Yukawa scheme  
 38 2               # 1- or 2-Loop RGEs 
 50 1               # Majorana phases: use only positive masses 
 51 0               # Write Output in CKM basis 
 52 1               # Write spectrum in case of tachyonic states 
 55 1               # Calculate one loop masses 
 57 0               # Calculate low energy constraints 
 65 1               # Solution tadpole equation 
\end{lstlisting}
\end{appendix}


\end{document}